\documentclass[10pt, letterpaper, onecolumn, draftcls]{IEEEtran}
\AtBeginDvi{}
\usepackage[cmex10]{amsmath}
\allowdisplaybreaks[1]
\usepackage{amssymb,amsthm}
\usepackage{mathrsfs}
\usepackage{cite}
\usepackage[dvipdf]{graphicx}
\usepackage{bm}

\theoremstyle{definition}
\newtheorem{theorem}{Theorem}
\newtheorem{lemma}{Lemma}
\newtheorem{corollary}{Corollary}
\newtheorem{remark}{Remark}
\newtheorem{definition}{Definition}

\newcommand{\eqtri}{\triangleq}

\newcommand{\bX}{\bm{X}}
\newcommand{\bY}{\bm{Y}}
\newcommand{\bx}{\bm{x}}
\newcommand{\by}{\bm{y}}
\newcommand{\cX}{\mathcal{X}}
\newcommand{\cY}{\mathcal{Y}}
\newcommand{\cA}{\mathcal{A}}
\newcommand{\cB}{\mathcal{B}}
\newcommand{\cC}{\mathcal{C}}
\newcommand{\cE}{\mathcal{E}}

\newcommand{\cS}{\mathcal{S}}
\newcommand{\tcS}{\tilde{\cS}}

\newcommand{\cT}{\mathcal{T}}
\newcommand{\cU}{\mathcal{U}}

\newcommand{\tx}{\tilde{x}}

\newcommand{\tQl}{\tilde{Q}^{(\rho)}}

\newcommand{\Ex}{\mathbb{E}}
\newcommand{\abs}[1]{\left\lvert#1\right\rvert}
\newcommand{\funcabs}[1]{\left\lVert#1\right\rVert}

\newcommand{\code}{\Phi}
\newcommand{\error}{p_e}

\newcommand{\expguess}{E_{\text{g}}}
\newcommand{\expsc}{E_{\text{s}}}

\title{On the Conditional Smooth R\'enyi Entropy and Its Applications in Guessing and Source Coding}
\author{Shigeaki Kuzuoka~\IEEEmembership{Member,~IEEE}
\thanks{The work was supported in part by JSPS KAKENHI Grant Number 18K04141.}
\thanks{S.~Kuzuoka is with the Faculty of Systems Engineering, Wakayama University, 930 Sakaedani, Wakayama, 640-8510 Japan, e-mail:kuzuoka@ieee.org.}
}

\begin{document}
\flushbottom
\maketitle

\begin{abstract} 
A novel definition of the conditional smooth R\'enyi entropy, which is
different from that of Renner and Wolf, is introduced.  It is shown that
our definition of the conditional smooth R\'enyi entropy is appropriate
to give lower and upper bounds on the optimal guessing moment in a
guessing problem where the guesser is allowed to stop guessing and
declare an error.  Further a general formula for the optimal guessing
exponent is given. In
particular, a single-letterized formula for mixture of i.i.d.~sources is
obtained. Another application in the problem of source coding with the
common side-information available at the encoder and decoder is also
demonstrated.
\end{abstract}

\begin{IEEEkeywords}
guessing,
information-spectrum method,
side information,
source coding, 
the conditional smooth R\'enyi entropy
\end{IEEEkeywords}

\section{Introduction}
Let us consider the problem of guessing the value of a random variable
$X$ by asking question of the form ``Is $X$ equal to $x$?''.  This
guessing game was introduced by Massey \cite{Massey94}, where the
average $\Ex[G(X)]$ of the number $G(x)$ of guesses required when $X=x$
was investigated. Subsequently Arikan \cite{Arikan96} gave a tight bound
on the guessing moment $\Ex[G(X)^\rho]$ for $\rho\geq 0$. He also investigated the guessing
problem of $X$ with the side-information $Y$. The result of Arikan
\cite{Arikan96} shows that the R\'enyi entropy \cite{Renyi61} (resp.~the conditional
R\'enyi entropy \cite{Arimoto77}) plays an important role to give upper and lower bounds
on $\Ex[G(X)^\rho]$ (resp.~$\Ex[G(X|Y)^\rho]$).

In this paper, we consider a variation of the problem of guessing $X$
with the side-information $Y$ such that the guesser is allowed to stop
guessing and declare an error.  We evaluate the expected value of the
cost of guessing under the constraint on the probability of the error.
To do this, we introduce the \emph{conditional smooth R\'enyi entropy}.

The concept of the ``smoothed'' version of the R\'enyi entropy was
introduced by Renner and Wolf \cite{RennerWolf_isit04,RennerWolf05}.
They defined the conditional $\varepsilon$-smooth R\'enyi entropy
$\tilde{H}_\alpha^\varepsilon(X|Y)$ of order $\alpha$, and showed the
significance of two special cases of $\tilde{H}_\alpha^\varepsilon(X|Y)$
in coding problems; Roughly speaking, they showed (i)
$\tilde{H}_0^\varepsilon(X|Y)\eqtri\lim_{\alpha\to
0}\tilde{H}_\alpha^\varepsilon(X|Y)$ characterizes the minimum codeword length in the source coding problem of $X$ with the
side-information $Y$ available at the decoder under the constraint that
the probability of decoding error is at most $\varepsilon$, and (ii)
$\tilde{H}_\infty^\varepsilon(X|Y)\eqtri\lim_{\alpha\to
\infty}\tilde{H}_\alpha^\varepsilon(X|Y)$ characterizes the amount of
uniform randomness that can be extracted from $X$.

Seeing the results of Arikan\cite{Arikan96} and Renner and Wolf
\cite{RennerWolf_isit04,RennerWolf05}, it is natural to expect that we
can use $\tilde{H}_\alpha^\varepsilon(X|Y)$ to characterize the cost of
guessing arrowing error.  However, the definition of
$\tilde{H}_\alpha^\varepsilon(X|Y)$ is not appropriate to be used in the
analysis of the guessing problem.  In this paper, we introduce another
``smoothed'' version $H_\alpha^\varepsilon(X|Y)$ of the conditional
R\'enyi entropy.  Then, by using $H_\alpha^\varepsilon(X|Y)$, we give
lower and upper bounds on the minimum cost of guessing the value of $X$
with the side-information $Y$ under the constraint that the guessing
error probability is at most $\varepsilon$.  Further, we demonstrate another
application of $H_\alpha^\varepsilon(X|Y)$ in the source coding problem.
Our contributions are summarized as follows.

\subsection{Contributions}
First we introduce a novel definition of the conditional
$\varepsilon$-smooth R\'enyi entropy $H_\alpha^\varepsilon(X|Y)$ of
order $\alpha$, and then investigate its properties.  In our definition,
similar to that of Renner and Wolf, the minimization over the set of
non-negative functions satisfying a particular condition is
involved. Our first contribution, Theorem \ref{prop:Q}, characterizes
the non-negative function $Q$ attaining the minimum in the definition of
$H_\alpha^\varepsilon(X|Y)$ for $\alpha\in (0,1)$.  This characterization is useful in the
proof of our theorems in guessing mentioned below.
Further, we investigate the asymptotic behavior of
$H_\alpha^\varepsilon(X|Y)$ by using the information spectrum method
\cite{Han-spectrum}.  Particularly, in Theorem \ref{thm:mixture}, we
show that the asymptotic value $H_\alpha^\varepsilon(\bX|\bY)$ of the
conditional smooth R\'enyi entropy for the mixture of i.i.d.~sources is
determined by the conditional entropy $H(X_i|Y_i)$ of a component of the
mixture.  This result allows as to give a singe-letterized formula in
guessing and source coding mentioned below.

Next we investigate the problem of ``guessing allowing error'', i.e.,
the problem of guessing $X$ with the side-information $Y$ where the
guesser can stochastically choose (i) to give up and declare an error or
(ii) to continue guessing at each step of guessing. The cost of guessing
is evaluated in the same way as Arikan \cite{Arikan96}; the cost is
$i^\rho$ for some $\rho>0$ if the value is correctly guessed at the
$i$-th step.  We consider the minimization of the expected value
$\bar{C}_\rho$ of the guessing cost under the the constraint that the
error probability $p_e$ is at most $\varepsilon$.  Our results, Theorems
\ref{thm:converse} and \ref{thm:direct}, give lower and upper bounds on
the minimum cost by using $H_{1/(1+\rho)}^\varepsilon(X|Y)$.  Further, a
general formula for the exponent of the optimal guessing cost is derived;
see Theorem \ref{thm:general_guess}.  \footnote{By general formula, we
mean that we consider sequences of guessing problems and do not place
any underlying structure such as stationarity, memorylessness and
ergodicity on the source \cite{Han-spectrum,ShunVincent2ndOrder}.}  In particular, a
single-letterized formula is given for mixture of i.i.d.~sources.
Moreover, our result for i.i.d.~sources demonstrates that allowing 
the vanishing error (i.e., $p_e\to 0$ as $n\to\infty$) drastically changes the problem from 
Arikan's original problem, where the \emph{zero-error} (i.e., $p_e=0$ for all $n$) is required.

The last contribution of this paper is to show the significance of our
conditional smooth R\'enyi entropy $H_\alpha^\varepsilon(X|Y)$ in the
problem of source coding.  We consider the variable-length lossless
coding problem of the source $X$ with the common side-information $Y$
available at the encoder and decoder.  We allow the decoder to make a
decoding error with probability at most $\varepsilon$.  Then, we
evaluate the exponential moment $M_\rho$ of the codeword length.  In a
similar manner as in the guessing problem, our results show that
$H_\alpha^\varepsilon(X|Y)$ can be used to characterize the minimum
value of $M_\rho$; Theorems \ref{thm:main_converse} and
\ref{thm:main_direct} give lower and upper bounds on the minimum value
of $M_\rho$ by using $H_{1/(1+\rho)}^\varepsilon(X|Y)$, and then Theorem
\ref{thm:gen} gives a general formula for the exponent of the minimum
value of $M_\rho$.

\subsection{Related Work}

As mentioned above, the concept of smooth R\'enyi entropy was first introduced by
Renner and Wolf \cite{RennerWolf_isit04,RennerWolf05}.  
Properties of the smooth R\'enyi entropy is
investigated by Koga \cite{Koga_itw13} by using majorization theory. As shown in Corollary \ref{corollary:Koga}, one of results in \cite{Koga_itw13} can be obtained as a corollary of our Theorem \ref{prop:Q}.

It is known that two special cases, $\alpha=0$ and $\alpha=\infty$, of
the smooth R\'enyi entropies have clear operational meaning respectively
in the fixed-length source coding
\cite{RennerWolf_isit04,RennerWolf05,Uyematsu10} and the intrinsic
randomness problem
\cite{RennerWolf_isit04,RennerWolf05,UyematsuKunimatsu_itw13}.
Similarly, the \emph{smooth R\'enyi divergence} also finds applications
in several coding problems; see, e.g.,
\cite{DattaRenner09,WangColbeckRenner_isit09,Warsi_itw13}.  To the
author's best knowledge, this paper first gives clear operational
meaning of the conditional smooth R\'enyi entropy of order
$\alpha\in(0,1)$ in guessing and source coding.

As mentioned above, Arikan \cite{Arikan96} showed the significance of
the R\'enyi entropy in the problem of guessing.  Recently, tighter
bounds on guessing moments was given by Sason and Verd\'u
\cite{SasonVerdu18jun}, where the R\'enyi entropy is also used.  The
guessing problem has been studied in various contexts such as the
problem of guessing subject to distortion
\cite{ArikanMerhav98,SaitoMatsushima18arXiv}, investigation of a
large deviation perspective of guessing
\cite{HanawalSundaresan11,ChristiansenDuffy13}, and
the guesswork in multi-user systems \cite{ChristiansenDuffyPinCalmonMedard15}, etc.; see,
e.g.,~\cite{SasonVerdu18jun} and references therein.

Campbell \cite{Campbell65} proposed the exponential moment of the
codeword length as an alternative to the average codeword length as a
criterion for variable-length lossless source coding, and gave upper and
lower bounds on the exponential moment in terms of the R\'enyi entropy.
\footnote{It should be mentioned here that a general problem for the
optimization of the exponential moment of a given cost function was
investigated by Merhav \cite{Merhav11,Merhav_arXiv2011}.}  On the other
hand, the problem of variable-length source coding \emph{allowing
errors} was investigated under the criterion of the average codeword
length by Koga and Yamamoto \cite{KogaYamamoto05} and Kostina \emph{et
al.}  \cite{KostinaPolyanskiyVerdu14isit14,KostinaPolyanskiyVerdu14}.
In \cite{ISIT2016}, the author gave a generalization of Campbell's
result to the case where the decoding error is allowed.  Recently, a
similar result without the prefix condition of codewords was given by
Sason and Verd\'u \cite{SasonVerdu18jun}.  Our results in Section
\ref{sec:sourcecoding} can be seen as a generalization of
\cite{ISIT2016}, since the result of \cite{ISIT2016} is obtained by
letting $\abs{\cY}=1$ in our results.

In this paper, two problems of guessing and source coding are
investigated.  The relations between the limiting guessing exponent and
the limiting exponent of the moment generating function of codeword
lengths in source coding was pointed out by Arikan and Merhav
\cite{ArikanMerhav98}; see also \cite{HanawalSundaresan11}.\footnote{More recently Beirami \textit{et al.} \cite{Beirami_etal_arXiv} showed an interesting connection between guessing and data compression from the geometric perspective.}
As expected, Theorems \ref{thm:general_guess} and \ref{thm:gen} below
reveal the equivalence between the optimal asymptotic exponent
$\expguess$ of guessing cost and that $\expsc$ of the exponential moment
of codeword lengths in source coding.

\subsection{Paper Organization}
The rest of the paper is organized as follows.
In Section \ref{sec:renyi_entropy}, the conditional $\varepsilon$-smooth
R\'enyi entropy $H_\alpha^\varepsilon(X|Y)$ of order $\alpha$ is
defined, and its properties are investigated.  The problems of guessing
and source coding are investigated in Section \ref{sec:guessing} and
Section \ref{sec:sourcecoding} respectively.  Concluding remarks and
directions for future work are provided in Section \ref{sec:conclusion}.
To ensure that the main ideas are seamlessly communicated in the main
text, we relegate all proofs to the appendices.

\section{Conditional Smooth R\'enyi Entropy}\label{sec:renyi_entropy}
Let $\cX$ and $\cY$ be finite or countably infinite sets.
For $\varepsilon\in[0,1)$ and a probability distribution $P_{XY}$ on $\cX\times\cY$, 
let $\cB^\varepsilon(P_{XY})$ be the set of
non-negative functions $Q$ with domain $\cX\times\cY$ such that $Q(x,y)\leq P_{XY}(x,y)$,
for all $(x,y)\in\cX\times\cY$, and $\sum_{x,y}Q(x,y)\geq 1-\varepsilon$. 
Then,
for $\alpha\in(0,1)\cup(1,\infty)$, the \emph{conditional $\varepsilon$-smooth 
R\'enyi entropy of order $\alpha$} is defined as\footnote{Throughout this paper, log denotes the natural logarithm.}
\begin{align}
 H_\alpha^\varepsilon(X|Y)\eqtri\frac{\alpha}{1-\alpha}\log r_\alpha^\varepsilon(X|Y)
\label{def:csre}
\end{align}
where
\begin{align}
r_\alpha^\varepsilon(X|Y)\eqtri
 \inf_{Q\in\cB^\varepsilon(P_{XY})}\sum_{y\in\cY}\left[
\sum_{x\in\cX}[Q(x,y)]^\alpha\right]^{1/\alpha}.
\end{align}
In the following, we assume that $0<\alpha<1$.\footnote{As seen below, 
the conditional $\varepsilon$-smooth
R\'enyi entropy of order $\alpha=1/(1+\rho)\in(0,1)$
plays an important role in guessing and source coding.}
Hence, $H_\alpha^\varepsilon(X|Y)$ can be rewritten as
\begin{align}
 H_\alpha^\varepsilon(X|Y)&=
\inf_{Q\in\cB^\varepsilon(P_{XY})}
\frac{\alpha}{1-\alpha}\log \sum_{y\in\cY}\left[
\sum_{x\in\cX}[Q(x,y)]^\alpha\right]^{1/\alpha}\\
&=
\inf_{Q\in\cB^\varepsilon(P_{XY})}
\frac{\alpha}{1-\alpha}\log \sum_{y\in\cY}P_Y(y)\left[
\sum_{x\in\cX}\left[\frac{Q(x,y)}{P_Y(y)}\right]^\alpha\right]^{1/\alpha}.
\label{eq:rewrite}
\end{align}
In the case of $\varepsilon=0$, $H_\alpha^\varepsilon(X|Y)$ is equivalent to the conditional R\'enyi entropy of order $\alpha$:\footnote{It is introduced by Arimoto \cite{Arimoto77}.}
\begin{align}
 H_\alpha^0(X|Y)&=
\frac{\alpha}{1-\alpha}\log \sum_{y\in\cY}\left[
\sum_{x\in\cX}[P_{XY}(x,y)]^\alpha\right]^{1/\alpha}
\end{align}
In the case of $\abs{\cY}=1$,  $H_\alpha^\varepsilon(X|Y)$ is equivalent to the
\emph{$\varepsilon$-smooth
R\'enyi entropy of order $\alpha$}, which is defined as 
\begin{align}
 H_\alpha^\varepsilon(X)&\eqtri\inf_{Q\in\cB^\varepsilon(P_{X})}
\frac{\alpha}{1-\alpha}\log \left[
\sum_{x\in\cX}[Q(x)]^\alpha\right]^{1/\alpha}\\
&=\inf_{Q\in\cB^\varepsilon(P_{X})}
\frac{1}{1-\alpha}\log 
\sum_{x\in\cX}[Q(x)]^\alpha
\end{align}
where $\cB^\varepsilon(P_X)$ is the set of
non-negative functions $Q$ with domain $\cX$ such that $Q(x)\leq P_X(x)$,
for all $x\in\cX$, and $\sum_{x\in\cX}Q(x)\geq 1-\varepsilon$.

Here, it should be emphasized that our definition \eqref{def:csre} of $H_\alpha^\varepsilon(X|Y)$ is slightly different from that of Renner and Wolf \cite{RennerWolf05}. In \cite{RennerWolf05} the conditional smooth R\'enyi entropy is defined as
\begin{align}
 \tilde{H}_\alpha^\varepsilon(X|Y)\eqtri\frac{1}{1-\alpha}\log \tilde{r}_\alpha^\varepsilon(X|Y)
\end{align}
where
\begin{align}
\tilde{r}_\alpha^\varepsilon(X|Y)\eqtri
 \inf_{Q\in\cB^\varepsilon(P_{XY})}\max_{\substack{y\in\cY:\\ P_Y(y)>0}}
\sum_{x\in\cX}\left[\frac{Q(x,y)}{P_Y(y)}\right]^\alpha.
\end{align}
To see the difference, rewrite $\tilde{r}_\alpha^\varepsilon(X|Y)$ for $0<\alpha<1$ as
\begin{align}
 \tilde{H}_\alpha^\varepsilon(X|Y)&=
\inf_{Q\in\cB^\varepsilon(P_{XY})}
\frac{1}{1-\alpha}\log 
\max_{\substack{y\in\cY:\\ P_Y(y)>0}}
\sum_{x\in\cX}\left[\frac{Q(x,y)}{P_Y(y)}\right]^\alpha\\
&=
\inf_{Q\in\cB^\varepsilon(P_{XY})}
\frac{\alpha}{1-\alpha}\log 
\max_{\substack{y\in\cY:\\ P_Y(y)>0}}\left[
\sum_{x\in\cX}\left[\frac{Q(x,y)}{P_Y(y)}\right]^\alpha\right]^{1/\alpha}.
\label{eq:RW}
\end{align}
Comparing \eqref{eq:RW} with \eqref{eq:rewrite}, we can see that the \emph{average} of $[\sum_x[Q(x,y)/P_Y(y)]^\alpha]^{1/\alpha}$ is taken in $H_\alpha^\varepsilon(X|Y)$, while the \emph{maximum} is taken in $\tilde{H}_\alpha^\varepsilon(X|Y)$. Hence it is apparent that
\begin{align}
 H_\alpha^\varepsilon(X|Y)\leq \tilde{H}_\alpha^\varepsilon(X|Y)
\end{align}
and the equality does not hold in general. 

\begin{remark}
\label{remark:difference}
The author thinks that the availability of the side-information $Y$ makes the difference between $H_\alpha^\varepsilon(X|Y)$ and $\tilde{H}_\alpha^\varepsilon(X|Y)$.
In the problem of guessing considered in Section \ref{sec:guessing}, the guesser can change the strategy according to the given $y\in\cY$.
Similarly, in the problem of source coding with \emph{common} side-information considered in Section \ref{sec:sourcecoding}, the encoder can choose the encoding function according to the given $y\in\cY$.
Hence, in these problems, ``the average with respect to $Y$'' has the significance in the coding theorems.
On the other hand, in the problems considered in \cite{RennerWolf05}, the encoder (or the extractor) cannot access to $Y$ and have to prepare for the worst case. Hence, ``the maximum with respect to $Y$'' has the significance in \cite{RennerWolf05}.
\end{remark}

Now, we show several properties of $H_\alpha^\varepsilon(X|Y)$.
First we investigate $Q\in\cB^\varepsilon(P_{XY})$ attaining the $\inf$ in the definition of $H_\alpha^\varepsilon(X|Y)$.
To do this, we introduce some notations.
For each $y\in\cY$ and $i=1,2,\dots$, let $x_y^i$ be the $i$-th probable $x\in\cX$ given $y$; i.e., $(x_y^i)_{i=1}^\infty$ is defined so that
\begin{align}
 P_{X|Y}(x_y^1|y)\geq P_{X|Y}(x_y^2|y)\geq P_{X|Y}(x_y^3|y)\geq \cdots.
\end{align}
Then, for each $y\in\cY$ and a given $\varepsilon_y$ satisfying $0\leq \varepsilon_y< 1$, let $i_y^*=i_y^*(\varepsilon_y)$ be the minimum integer such that
\begin{align}
 \sum_{i=1}^{i_y^*}P_{X|Y}(x_y^i|y)\geq 1-\varepsilon_y
\label{eq:def:istar}
\end{align}
and let
\begin{align}
 Q_{\varepsilon_y}^*(x_y^i|y)\eqtri
\begin{cases}
 P_{X|Y}(x_y^i|y), & i=1,2,\dots, i_y^*-1,\\
 1-\varepsilon_y-\sum_{i=1}^{i^*-1}P_{X|Y}(x_y^i|y), & i=i_y^*,\\
 0, & i> i_y^*.
\end{cases}
\label{eq:def:Qstar}
\end{align}
For $\varepsilon_y=1$, let $i_y^*(1)=\infty$ and $Q_1^*(x|y)=P_{X|Y}(x|y)$ for all $(x,y)\in\cX\times\cY$.

\begin{theorem}
\label{prop:Q}
By using notations introduced above, we have
\begin{align}
 H_\alpha^\varepsilon(X|Y)&=
\inf_{(\varepsilon_y)_{y\in\cY}\in\cE_0(\varepsilon)}
\frac{\alpha}{1-\alpha}\log \sum_{y\in\cY}P_Y(y)\left[
\sum_{i=1}^{i_y^*}\left[Q_{\varepsilon_y}^*(x_y^i|y)\right]^\alpha\right]^{1/\alpha}
\label{eq:prop:Q}
\end{align}
where $\cE_0(\varepsilon)$ is the set of $(\varepsilon_y)_{y\in\cY}$ satisfying $0\leq \varepsilon_y\leq 1$, for all $y\in\cY$, and 
$\sum_y\varepsilon_yP_Y(y)=\varepsilon$.
\end{theorem}

Theorem \ref{prop:Q} will be proved in Appendix \ref{sec:proof_prop_Q}.
As a corollary, we have a known property of $H_\alpha^\varepsilon(X)$, which is proved in (A) of Theorem 1 of \cite{Koga_itw13}.

\begin{corollary}
\label{corollary:Koga}
 Assume that $x^1,x^2,\dots,$ are sorted so that $P_X(x^1)\geq P_X(x^2)\geq\dots$ and let $i^*$ be 
the minimum integer such that $\sum_{i=1}^{i^*}P_{X}(x^i)\geq 1-\varepsilon$.
Then 
\begin{align}
 Q_{\varepsilon}^*(x^i)\eqtri
\begin{cases}
 P_{X}(x^i), & i=1,2,\dots, i^*-1,\\
 1-\varepsilon-\sum_{i=1}^{i^*-1}P_{X|Y}(x^i), & i=i^*,\\
 0, & i> i^*
\end{cases}
\end{align}
attains the inifimum in the definition of $H_\alpha^\varepsilon(X)$; i.e., 
\begin{align}
 H_\alpha^\varepsilon(X)&=\frac{1}{1-\alpha}\log \sum_{i=1}^{i^*}\left[Q_{\varepsilon}^*(x^i)\right]^\alpha.
\end{align}
\end{corollary}

\medskip
Next, we investigate the asymptotic behavior of the conditional $\varepsilon$-smooth R\'enyi entropy by using the information spectrum method \cite{Han-spectrum}. Let us consider a pair of correlated general sources $(\bX,\bY)=\{(X^n,Y^n)\}_{n=1}^\infty$, which 
is a sequence of pairs $(X^n,Y^n)$ of correlated random variables $X^n$ on the $n$-th Cartesian product $\cX^n$ of $\cX$ and $Y^n$ on $\cY^n$.
The joint distribution of $(X^n,Y^n)$ is denoted by $P_{X^nY^n}$, which is not required to satisfy the consistency condition.
Given $(\bX,\bY)$, $\alpha\in(0,1)$, and $\varepsilon\in[0,1)$, let us define $H_{\alpha}^\varepsilon(\bX|\bY)$ as
\begin{align}
 H_{\alpha}^\varepsilon(\bX|\bY)\eqtri\lim_{\delta\downarrow 0}\limsup_{n\to\infty}\frac{1}{n}H_{\alpha}^{\varepsilon+\delta}(X^n|Y^n).
\label{def:CSRE}
\end{align}
As shown in the following sections, this quantity plays an important role in the general formulas of guessing and source coding.

Here it is worth to note that $H_{\alpha}^\varepsilon(\bX|\bY)$ is non-negative for all 
$\alpha\in(0,1)$ and $\varepsilon\in[0,1)$.
Indeed, we can prove the stronger fact that
\begin{align}
 \liminf_{n\to\infty}\frac{1}{n}H_{\alpha}^{\varepsilon}(X^n|Y^n)\geq 0,\quad \alpha\in(0,1),\varepsilon\in[0,1).
\label{eq:nonnegativity}
\end{align}
We will prove \eqref{eq:nonnegativity} in Appendix \ref{sec:proof_eq_nonnegativity}.

To give a single-letterized form of $H_{\alpha}^\varepsilon(\bX|\bY)$, we consider a mixture of i.i.d.~sources.
Let us consider $m$ distributions $P_{X_iY_i}$ ($i=1,2,\dots,m$) on $\cX\times\cY$. 
A general source $(\bX,\bY)$ is said to be a mixture of $P_{X_1Y_1},P_{X_2Y_2},\dots,P_{X_mY_m}$ if 
there exists $(\alpha_1,\alpha_2\dots,\alpha_m)$ satisfying (i) $\sum_i\alpha_i=1$, (ii) $\alpha_i>0$ ($i=1,\dots,m$), and (ii) for all $n=1,2,\dots$, all $x^n=(x_1,x_2,\dots,x_n)\in\cX^n$ and $y^n=(y_1,y_2,\dots,y_n)\in\cY^n$,
\begin{align}
 P_{X^nY^n}(x^n,y^n)&=\sum_{i=1}^m\alpha_iP_{X_i^nY_i^n}(x^n,y^n)\\
&=\sum_{i=1}^m\alpha_i\prod_{t=1}^nP_{X_iY_i}(x_t,y_t).
\label{eq:mix_iid}
\end{align}
For the later use, let $A_i\eqtri\sum_{j=1}^{i-1}\alpha_i$ ($i=1,2,\dots,m$) and $A_{m+1}\eqtri 1$.
Further, to simplify the analysis, we assume that
\begin{align}
 H(X_1|Y_1)> H(X_2|Y_2)>\dots>  H(X_m|Y_m)
\label{eq:common_sort}
\end{align}
where
$H(X_i|Y_i)$ is the conditional entropy determined by $P_{X_iY_i}$:
\begin{align}
 H(X_i|Y_i)\eqtri\sum_{x\in\cX}P_{X_iY_i}(x,y)\log\frac{1}{P_{X_i|Y_i}(x|y)}.
\end{align}
Then, $H_{\alpha}^{\varepsilon}(\bX|\bY)$ of the mixture $(\bX,\bY)$ is characterized as in the following theorem.

\begin{theorem}
\label{thm:mixture}  
Let $(\bX,\bY)$ be a mixture of i.i.d.~sources satisfying \eqref{eq:common_sort}.
Fix $\alpha\in(0,1)$, $i$, and $\varepsilon\in[A_i,A_{i+1})$. Then, we have 
\begin{align}
 H_{\alpha}^{\varepsilon}(\bX|\bX) =H(X_i|Y_i).
\label{eq:thm_mixture}
\end{align}
\end{theorem}

Theorem \ref{thm:mixture} will be proved in Appendix \ref{sec:proof_thm_mixture}.

\begin{remark}
Although Theorem \ref{thm:mixture} assumes that components are i.i.d., this assumption is not crucial.
Indeed, the property of i.i.d.~sources used in the proof of the theorem is only that the AEP \cite{Cover2} holds, i.e.,
\begin{align}
 \lim_{n\to\infty}\Pr\left\{
 \abs{
 \frac{1}{n}\log\frac{1}{P_{X_i^n}(X_i^n|Y_i^n)}-H(X_i|Y_i)} >\zeta
 \right\}=0
\end{align}
for all $i=1,2,\dots,m$ and any $\zeta>0$. 
Hence, it is straightforward to extend the theorem so that it can be applied for the mixture of stationary and ergodic sources.
\end{remark}

\section{Guessing}\label{sec:guessing}
In this section, we assume that the alphabet $\cX$ is finite; we assume that $\abs{\cX}=K$ and $\cX=\{1,2,\dots,K\}$.

A \emph{guessing strategy} $G=((\sigma_y,\pi_y))_{y\in\cY}$ for $X$ given $Y$ is defined by a collection of 
pairs $(\sigma_y,\pi_y)$, for each $y$, of (i) a permutation $\sigma_y$ on $\cX$ and (ii) a 
vector $\pi_y=(\pi_y(i))_{i=1}^K$ satisfying $0\leq\pi_y(i)\leq 1$ for
all $i=1,2,\dots,K$.
Given the side information $y\in\cY$, the ``guesser'' corresponding to the strategy $G$ guesses the value of $X$ as the following manner.  At the $i$th step
($i=1,2,\dots,K$), the guesser determines whether to ``give up'' or not;
the guesser gives up and stops guessing with probability $\pi_y(i)$ and the error of guessing is declared. If
the guesser does not give up then the value $x\in\cX$ satisfying $\sigma_y(x)=i$ is chosen as the
``guessed value''.  The guessing will be continued until when the
guesser gives up or when the value of $X$ is correctly guessed (i.e.,
$\sigma_y^{-1}(i)=X$ at the $i$th step). It should be noted here that the guessing function studied in \cite{Arikan96} corresponds to the guessing strategy such that $\pi_y(i)=0$ for all $y\in\cY$ and $i=1,2,\dots,K$.

In this paper, we evaluate the ``cost'' of guessing as follows.
If the guessing is stopped before the value of $X$ is correctly guessed then a constant cost $c_e\geq 0$ is given as ``penalty''.
Otherwise, the cost of guessing is given by $i^\rho$ when the value of $X$ is correctly guessed at the $i$th step, where $\rho\geq 0$ is a constant.
For each $y\in\cY$, let
\begin{align}
 \lambda_y(i)\eqtri\prod_{j=1}^i(1-\pi_y(j)),\quad i=1,2,\dots,K.
\label{eq:zeta}
\end{align}
Then we can see that, given $y\in\cY$, the conditional probability of the event ``the value of $X$ is correctly guessed at the $i$-th step before give up'' is
\begin{align}
 \lambda_y(i)P_{X|Y}(\sigma_y^{-1}(i)|y)
\end{align}
and thus, the conditional probability of the event ``the guesser gives up before guessing the the value of $X$ correctly'' is
\begin{align}
\lefteqn{1-\sum_{i=1}^K\lambda_y(i)P_{X|Y}(\sigma_y^{-1}(i)|y)}\nonumber\\
&=1-\sum_{x\in\cX}\lambda_y(\sigma_y(x))P_{X|Y}(x|y).
\end{align}
Hence, the error probability $p_e=p_e(G|X,Y)$, i.e., the average probability such that the guessing is stopped before the value of $X$ is correctly guessed, is given by
\begin{align}
 p_e&=\sum_{y\in\cY}P_Y(y)\left\{
1-\sum_{i=1}^K\lambda_y(i)P_{X|Y}(\sigma_y^{-1}(i)|y)
\right\}\\
&=\sum_{y\in\cY}P_Y(y)\left\{
1-\sum_{x\in\cX}\lambda_y(\sigma_y(i))P_{X|Y}(x|y)
\right\}\\
&=
1-\sum_{(x,y)\in\cX\times\cY}\lambda_y(\sigma_y(x))P_{XY}(x,y)
,\label{eq:p_e}
\end{align}
and the expected value $\bar{C}_\rho'=\bar{C}_\rho'(G|X,Y)$ of the cost is given by
\begin{align}
 \bar{C}_\rho'&=\sum_{y\in\cY}P_Y(y)\left\{
\sum_{i=1}^K\lambda_y(i)P_{X|Y}(\sigma_y^{-1}(i)|y)i^\rho
\right\}+p_ec_e.
\label{eq:barCprime}
\end{align}

For some applications, it may be natural to simply minimize the cost $\bar{C}_\rho'$.
However, for some applications, it is not easy to evaluate the precise value $c_e$ of the penalty for stopping the guessing.\footnote{Although we assume that $c_e$ is a constant, it may depend on the true value of $X$, the number of guesses before giving up, etc; It heavily depends on the application. Hence we leave the general cost case as a future work, and concentrate on the first term of \eqref{eq:barCprime}.}
In such a situation, we may consider the cost of guessing and the penalty separately, and 
minimize
\begin{align}
 \bar{C}_\rho(G|X,Y)&\eqtri\sum_{y\in\cY}P_Y(y)\left\{
\sum_{i=1}^K\lambda_y(i)P_{X|Y}(\sigma_y^{-1}(i)|y)i^\rho
\right\}\\
&=
\sum_{y\in\cY}P_Y(y)\left\{
\sum_{x\in\cX}\lambda_y(\sigma_y(x))P_{X|Y}(x|y)\sigma_y(x)^\rho\right\}\\
&=
\sum_{(x,y)\in\cX\times\cY}\lambda_y(\sigma_y(x))P_{XY}(x,y)\sigma_y(x)^\rho
\label{eq:barC}
\end{align}
under the constraint on the probability $p_e$ of stopping the guessing. 
Further, if the minimum value $\bar{C}_\rho^*(\varepsilon)$ of $\bar{C}_\rho$ under the constraint that $p_e\leq\varepsilon$ is known,
it is not hard to optimize $\bar{C}_\rho'$; the optimal value can be written as
$\inf_{\varepsilon}[\bar{C}_\rho^*(\varepsilon)+\varepsilon c_e]$.
So, we study the minimizing problem of $\bar{C}_\rho$ under the constraint that $p_e\leq\varepsilon$ for a given constant $0\leq\varepsilon\leq 1$; the results are summarized in the following theorems:

\begin{theorem} 
\label{thm:converse}
Fix $\rho>0$ and $\varepsilon\in[0,1)$.
For any guessing strategy $G$ satisfying $p_e(G|X,Y)\leq\varepsilon$, the expected value $\bar{C}_\rho=\bar{C}_\rho(G|X,Y)$ of the cost must satisfy
\begin{align}
\bar{C}_\rho\geq (1+\log K)^{-\rho}\exp\left\{\rho H_{\frac{1}{1+\rho}}^\varepsilon(X|Y)\right\}. 
\label{eq:converse}
\end{align}
\end{theorem}

\begin{theorem}
\label{thm:direct}
Fix $\rho>0$ and $\varepsilon\in[0,1)$.
There exists a guessing strategy $G$ such that the error probability satisfies $p_e(G|X,Y)\leq\varepsilon$ and the expected value $\bar{C}_\rho=\bar{C}_\rho(G|X,Y)$ of the cost satisfies
\begin{align}
\bar{C}_\rho\leq \exp\left\{\rho H_{\frac{1}{1+\rho}}^\varepsilon(X|Y)\right\}. 
\label{eq:direct}
\end{align}
\end{theorem}

\medskip
Theorems \ref{thm:converse} and \ref{thm:direct} will be proved in Appendix \ref{sec:proof}.

\begin{remark}
\label{remark:dice}
The proof of Theorem \ref{thm:direct} reveals the fact that 
``the optimal guesser throws a dice at most once''; i.e., for all $y\in\cY$, $\pi_y=(\pi_y(i))_{i=1}^K$ of the optimal strategy $G$ satisfies
\begin{align}
 \pi_y(i)=
\begin{cases}
 0 &: i< i_y^*\\
 1 &: i> i_y^*
\end{cases}
\end{align}
and $0\leq\pi_y(i^*)\leq 1$ for some $i^*$ ($i^*=0,1,\dots,K$).
In other words, the optimal guesser makes guesses $i^*-1$ times (or until the value of $X$ is correctly guessed), and then, 
 moves on the $i^*$th guessing with the probability $1-\pi_y(i^*)$.
\end{remark}

\medskip
Now, let us consider the asymptotic behavior of the cost of guessing. 
Particularly we investigate the asymptotic behavior of the exponent of the cost.
So, we define the achievability of the exponential value as follows.

\begin{definition}
Given a constant $\rho>0$ and a general source $(\bX,\bY)$, a value $\expguess$ is said to be $\varepsilon$-achievable if there exists a sequence  $\{G_n\}_{n=1}^\infty$ of strategies satisfying
\begin{align}
 \limsup_{n\to\infty}p_e(G_n|X^n,Y^n)\leq\varepsilon
\end{align}
and
\begin{align}
 \limsup_{n\to\infty}\frac{1}{n}\log\bar{C}_\rho(G_n|X^n,Y^n)\leq \expguess.
\end{align}
The infimum of $\varepsilon$-achievable values is denoted by $\expguess(\rho,\varepsilon|\bX,\bY)$.
\end{definition}

\medskip
Then we have the following theorem, where $H_{\alpha}^\varepsilon(\bX|\bY)$ defined in \eqref{def:CSRE} characterized the optimal asymptotic exponent of the guessing cost.

\begin{theorem}
\label{thm:general_guess}
For any $\rho>0$ and $\varepsilon\in[0,1)$,
 \begin{align}
  \expguess(\rho,\varepsilon|\bX,\bY)=\rho H_{1/(1+\rho)}^\varepsilon(\bX|\bY).
 \end{align}
\end{theorem}

The theorem will be proved in Appendix \ref{sec:proof_general_guess}.
Combining Theorem \ref{thm:general_guess} with Theorem \ref{thm:mixture}, we can obtain the single-letterized characterization of $\expguess(\rho,\varepsilon|\bX,\bY)$ for a mixed source $(\bX,\bY)$.

\begin{corollary}
\label{corollary:guess_mix}
Let $(\bX,\bY)$ be a mixture of i.i.d.~sources satisfying \eqref{eq:common_sort}.
Then, for any $\rho>0$ and $\varepsilon\in[0,1)$,
\begin{align}
 \expguess(\rho,\varepsilon|\bX,\bY) =\rho H(X_i|Y_i)
\end{align}
where $i$ is determined so that $\varepsilon\in[A_i,A_{i+1})$.
\end{corollary}

In particular, let us consider a special case of guessing for an i.i.d.~source (i.e., $m=1$) under the constraint $\varepsilon=0$.
Corollary \ref{corollary:guess_mix} shows 
\begin{align}
 \expguess(\rho,0|\bX,\bY) =\rho H(X|Y).
\end{align}
In other words, $\expguess(\rho,0|\bX,\bY)$ is determined by the
parameter $\rho$ and the conditional entropy $H(X|Y)$ of the source.  On
the other hand, Proposition 5 of \cite{Arikan96} shows that the exponent
of the optimal guessing for an i.i.d.~source is $\rho
H_{1/(1+\rho)}^0(X|Y)$, which is characterized by $\rho$ and the
conventional R\'enyi entropy.  It may seem to be a contradiction, but it
is not.  The constraint $\varepsilon=0$ in our problem requires that
$p_e\to 0$ as $n\to\infty$; i.e., the vanishing error is allowed.  On
the other hand, Arikan's original guessing problem \cite{Arikan96}
requires the \emph{zero-error}; i.e., it is required that $p_e=0$ for
all $n$.  Our result shows that allowing the vanishing error changes the
problem drastically.

\begin{remark}
 It is well known that, in the channel coding problem, the
 zero-error capacity \cite{Shannon56} is quite different from the
 conventional capacity. The above argument demonstrates that an analogue
 holds also in the guessing problem.
\end{remark}

\section{Source Coding}\label{sec:sourcecoding}

A variable-length source code $\code=\left((\varphi_y,\psi_y,\cC_y)\right)_{y\in\cY}$ for $X$ with the common-side-information $Y$ is determined a collection of triplets $(\varphi_y,\psi_y,\cC_y)$, for each $y\in\cY$, of (i) a set $\cC_y\subset\{0,1\}^*$ of finite-length binary strings, (ii) an stochastic encoder mapping $\varphi_y\colon\cX\to\cC$, and (iii) a decoder mapping $\psi_y\colon\cC\to\cX$.
Without loss of generality, we assume that $\cC_y=\{\varphi_y(x):x\in\cX\}$. Further, we assume that $\cC_y$ satisfies the prefix condition for each $y\in\cY$.

Note that we allow the encoder mapping $\varphi_y$ to be stochastic. 
Let $W_{\varphi_y}(c|x)$ be the probability that $x\in\cX$ is encoded in $c\in\cC_y$ by $\varphi_y$. Then,
the error probability $\error=\error(\code|X,Y)$ of the code $\code$ is defined as
\begin{align}
 \error&\eqtri\sum_{y\in\cY}P_Y(y)\Pr\left\{X\neq\psi_y(\varphi(X))\right\}\\
&=\sum_{y\in\cY}P_Y(y)\left\{
\sum_{x\in\cX}P_{X|Y}(x|y)\sum_{c:x\neq\psi_y(c)}W_{\varphi_y}(c|x)
\right\}\\
&=\sum_{(x,y)\in\cX\times}P_{XY}(x,y)\sum_{c:x\neq\psi_y(c)}W_{\varphi_y}(c|x).
\end{align}

The length of the codeword $\varphi_y(x)$ of $x$ (in bits) is denoted by $\funcabs{\varphi_y(x)}$. Let $\ell(\cdot|y)$ be the length function (in nats):
\begin{align}
 \ell(x|y)\eqtri\funcabs{\varphi_y(x)}\log 2.
\end{align}
In this study, we focus on the exponential moment of the length function. 
For a given $\rho>0$, let us define the exponential moment $M_\rho=M_\rho(\code|X,Y)$ of the length function as
\begin{align}
 M_\rho&\eqtri\Ex_{P_{XY}}\left[\exp\{\rho\ell(X|Y)\}\right]\\
&=\sum_{y\in\cY}P_Y(y)\left\{
\sum_{x\in\cX}P_{X|Y}(x|y)\sum_{c\in\cC_y}W_{\varphi_y}(c|x)\exp\{\rho\funcabs{c}\log 2\}
\right\}\\
&=
\sum_{(x,y)\in\cX\times\cY}P_{XY}(x,y)\sum_{c\in\cC_y}W_{\varphi_y}(c|x)\exp\{\rho\funcabs{c}\log 2\}.
\label{eq:def_moment}
\end{align}
subject to $\error(\code)\leq\varepsilon$, where $\Ex_P$ denotes the expectation with respect to the distribution $P$.

\begin{remark}
Without loss of optimality we can assume that the decoder mapping $\psi_y$ is deterministic for all $y\in\cY$. Indeed, for a given $W_{\varphi_y}$, we can choose $\psi_y$ so that
\begin{align}
 \psi_y(c)=\arg\max_{x\in\cX} W_{\varphi_y}(c|x)P_{X|Y}(x|y).
\end{align}
\end{remark}

\medskip
We consider the problem of minimizing $M_\rho$ under the constraint that $p_e\leq\varepsilon$ for a given constant $0\leq\varepsilon\leq 1$; the results are summarized in the following theorems:

\begin{theorem}
\label{thm:main_converse} 
Fix $\rho>0$ and $\varepsilon\in[0,1)$. For any code $\code$ satisfying $\error(\code|X,Y)\leq\varepsilon$, the moment $M_\rho=M_\rho(\code|X,Y)$ must satisfy
\begin{align}
M_\rho\geq 
\exp\left\{\rho H_{\frac{1}{1+\rho}}^\varepsilon(X|Y)\right\}.
\label{eq:thm_main_converse}
\end{align}
\end{theorem}

\begin{theorem}
\label{thm:main_direct} 
Fix $\rho>0$ and $\varepsilon\in[0,1)$. There exists a code $\code$ such that $\error(\code|X,Y)\leq\varepsilon$ and $M_\rho=M_\rho(\code|X,Y)$ satisfies
\begin{align}
M_\rho
\leq 
2^{2\rho}\exp\left\{\rho H_{\frac{1}{1+\rho}}^\varepsilon(X|Y)\right\}+\varepsilon 2^{\rho}.
\label{eq:thm_main_direct}
\end{align}
\end{theorem}

\medskip
Theorems \ref{thm:main_converse} and \ref{thm:main_direct} and  will be proved in Appendix \ref{sec:proof_sourcecoding}.

\begin{remark}
While we allow the encoder mapping $\varphi$ to be stochastic in Theorem
\ref{thm:main_direct}, we can see the fact that ``the optimal encode
throws a dice at most once''; cf.~Remark \ref{remark:dice}.  Hence, it
is not hard to modify the theorem for the case where only deterministic
encoder mappings are allowed.
We omit the details, but see Proposition 1 of \cite{ISIT2016} for the case of $\abs{\cY}=1$.
\end{remark}

\medskip
Now, let us consider the asymptotic behavior of the exponential moment of the length function. 
In a same way as $\expguess$, we define the achievability of the exponential value as follows.

\begin{definition}
Given a constant $\rho>0$ and a general source $(\bX,\bY)$, a value $\expsc$ is said to be $\varepsilon$-achievable if there exists a sequence  $\{\code_n\}_{n=1}^\infty$ of variable-length codes satisfying
\begin{align}
 \limsup_{n\to\infty}p_e(\code_n|X^n,Y^n)\leq\varepsilon
\end{align}
and
\begin{align}
 \limsup_{n\to\infty}\frac{1}{n}\log M_\rho(\code_n|X^n,Y^n)\leq \expsc.
\end{align}
The infimum of $\varepsilon$-achievable values is denoted by $\expsc(\rho,\varepsilon|\bX,\bY)$.
\end{definition}
\medskip

Then we have the following general formula, which will be proved in Appendix \ref{sec:proof_thm_gen}.

\begin{theorem}
\label{thm:gen}
For any $\rho>0$ and $\varepsilon\in[0,1)$,
\begin{align}
 \expsc(\rho,\varepsilon|\bX,\bY)=\rho H_{1/(1+\rho)}^{\varepsilon}(\bX|\bY).
\end{align}
\end{theorem}
\medskip

Combining Theorem \ref{thm:gen} with Theorem \ref{thm:mixture}, we can obtain the single-letterized characterization of $\expsc(\rho,\varepsilon|\bX,\bY)$ for a mixed source $(\bX,\bY)$.

\begin{corollary}
Let $(\bX,\bY)$ be a mixture of i.i.d.~sources satisfying \eqref{eq:common_sort}.
Then, for any $\rho>0$ and $\varepsilon\in[0,1)$,
\begin{align}
 \expsc(\rho,\varepsilon|\bX,\bY) =\rho H(X_i|Y_i)
\end{align}
where $i$ is determined so that $\varepsilon\in[A_i,A_{i+1})$.
\end{corollary}

\section{Concluding Remarks}\label{sec:conclusion}

In this paper, a novel definition of the conditional smooth R\'enyi
entropy was introduced, and its significance in the problems of guessing
and source coding was demonstrated.

Although properties of $H_\alpha^\varepsilon(X|Y)$ and
$H_\alpha^\varepsilon(\bX|\bY)$ are investigated in Section \ref{sec:renyi_entropy}, we consider only the
case of $\alpha\in(0,1)$.  It is an important future work to investigate
the properties of $H_\alpha^\varepsilon(X|Y)$ and
$H_\alpha^\varepsilon(\bX|\bY)$ of order $\alpha>1$.  On the other hand,
in the coding theorems in Sections \ref{sec:guessing} and
\ref{sec:sourcecoding}, it is sufficient to consider the the conditional
smooth R\'enyi entropy of order $\alpha=1/(1+\rho)\in(0,1)$.  It is an
important future work to find the operational meaning of
$H_\alpha^\varepsilon(X|Y)$ of order $\alpha>1$.

In Section \ref{sec:sourcecoding}, we assume that the common side-information $Y$ is available at encoder and decoder.
As mentioned in Remark \ref{remark:difference}, the availability of $Y$ at the encoder is very important.
The author conjectures that, in the case such that $Y$ is available only at decoder, $\tilde{H}_\alpha^\varepsilon(X|Y)$ instead of $H_\alpha^\varepsilon(X|Y)$ characterizes the exponential moment of codeword lengths. We leave it as a future work.

\appendices
\section{Proof of Theorem \ref{prop:Q}}\label{sec:proof_prop_Q}

Before proving the theorem, we give several lemmas in Subsection \ref{sec:proof_prop_Q_Lemmas}. The proof of the theorem is given in Subsection \ref{sec:proof_prop_Q_main}.

\subsection{Lemmas}\label{sec:proof_prop_Q_Lemmas}

\begin{lemma}
\label{lemma:Schur}
Let $k$ be a finite integer. Assume that $\bm{p}=(p_i)_{i=1}^k$ and $\bm{q}=(q_i)_{i=1}^k$ satisfy $p_i\geq 0$, $q_i\geq 0$ ($i=1,2,\dots,k$),
\begin{align}
 \sum_{i=1}^jp_i\leq \sum_{i=1}^jq_i,\quad j=1,2,\dots,k-1,
\end{align}
and
\begin{align}
 \sum_{i=1}^np_i= \sum_{i=1}^nq_i.
\end{align}
Then, for all $\alpha\in(0,1)$, we have
\begin{align}
 \sum_{i=1}^kp_i^\alpha\geq \sum_{i=1}^kq_i^\alpha.
\end{align}
\end{lemma}

The lemma is a consequence of the \emph{Schur concavity} of the function $f(\bm{p})=\sum_{i=1}^k p_i^\alpha$ for $\alpha\in(0,1)$; 
$f$ is Schur concave and the assumption of the lemma means $\bm{p}\prec\bm{q}$ (i.e., $\bm{q}$ majorizes $\bm{p}$), hence we have $f(\bm{p})\geq f(\bm{q})$. See \cite{Marshall1979Inequalities} for more details. 

By modifying Lemma \ref{lemma:Schur}, we have the following key lemma.

\begin{lemma}
\label{lemma:Schur2}
Fix $\alpha\in(0,1)$, $\varepsilon\in(0,1]$ and $\bm{p}=(p_i)_{i=1}^\infty$ satisfying (i) $p_i\geq 0$ for all $i=1,2,\dots$, (ii) $p_1\geq p_2\geq p_3\geq\cdots$,  and (iii) $\sum_ip_i=1$.
Let $i^*$ be the minimum integer such that $\sum_{i=1}^{i^*}p_i\geq 1-\varepsilon$ and let
\begin{align}
 q_i^*\eqtri
\begin{cases}
 p_i, & i=1,2,\dots, i^*-1,\\
 1-\varepsilon-\sum_{i=1}^{i^*-1}p_i, & i=i^*,\\
 0, & i> i^*.
\end{cases}
\end{align}
Then we have
\begin{align}
\sum_{i=1}^\infty (q_i^*)^\alpha = \inf_{\bm{q}\in\cB^\varepsilon(\bm{p})}\sum_{i=1}^\infty q_i^\alpha
\end{align}
where $\cB^\varepsilon(\bm{p})$ is the set of $\bm{q}=(q_i)_{i=1}^\infty$
satisfying $0\leq q_i\leq p_i$ ($i=1,2,\dots$) and $\sum_iq_i\geq 1-\varepsilon$.
\end{lemma}

\begin{IEEEproof}
Let
\begin{align}
 \cB_0^\varepsilon\eqtri\left\{\bm{q}\in\cB^\varepsilon(\bm{p}):\sum_iq_i= 1-\varepsilon\right\}
\end{align}
and
\begin{align}
 \cB_{0,\text{finite}}^\varepsilon\eqtri\left\{\bm{q}\in\cB_0^\varepsilon(\bm{p}):\exists k\text{ s.t. }q_i=0\text{ for all }i>k\right\}.
\end{align}
For any $\bm{q}\in\cB^\varepsilon(\bm{p})$, there exists $\hat{\bm{q}}\in\cB_0^\varepsilon(\bm{p})$ satisfying $0\leq \hat{q}_i<q_i$ for all $i=1,2,\dots$. Since $\sum_i\hat{q}_i^\alpha\leq \sum_iq_i^\alpha$, we have
\begin{align}
\inf_{\bm{q}\in\cB_0^\varepsilon(\bm{p})}\sum_{i=1}^\infty q_i^\alpha
=\inf_{\bm{q}\in\cB^\varepsilon(\bm{p})}\sum_{i=1}^\infty q_i^\alpha.
\end{align}
On the other hand, Lemma \ref{lemma:Schur} guarantees that
\begin{align}
\sum_{i=1}^\infty (q_i^*)^\alpha = \inf_{\bm{q}\in\cB_{0,\text{finite}}^\varepsilon(\bm{p})}\sum_{i=1}^\infty q_i^\alpha.
\end{align}

So, to prove the lemma, it is sufficient to prove the following fact: For any $\bm{q}$ satisfying
$\bm{q}\in\cB_0^\varepsilon(\bm{p})$ and $\bm{q}\notin\cB_{0,\text{finite}}^\varepsilon(\bm{p})$, there exists
$\hat{\bm{q}}\in\cB_{0,\text{finite}}^\varepsilon(\bm{p})$ such that $\sum_i\hat{q}_i^\alpha\leq \sum_iq_i^\alpha$.
Indeed, the fact can be proved as follows.
Since $\bm{q}\in\cB_0^\varepsilon(\bm{p})$ and $\bm{q}\notin\cB_{0,\text{finite}}^\varepsilon(\bm{p})$, we can choose finite integers $j$ and $k$ such that $q_j<p_j$, $k>j$, and $\sum_{i=k+1}^\infty q_i<p_j-q_j$. Then, choose $\hat{\bm{q}}$ so that
\begin{align}
 \hat{q}_i\eqtri
\begin{cases}
 q_i, & i=1,2,\dots,j-1,j+1,\dots,k\\
 q_j+\sum_{i'=k+1}^\infty q_{i'}, & i=j,\\
 0, & i> k.
\end{cases}
\end{align}
Since $t^\alpha\geq t$ for $0\leq t\leq 1$, we have
\begin{align}
 \left(\frac{q_j}{\hat{q}_j}\right)^\alpha
+\sum_{i'=k+1}^\infty  \left(\frac{q_{i'}}{\hat{q}_j}\right)^\alpha
&\geq \frac{q_j}{\hat{q}_j}+\sum_{i'=k+1}^\infty  \frac{q_{i'}}{\hat{q}_j} =1
\end{align}
and thus
\begin{align}
 q_j^\alpha+\sum_{i'=k+1}^{\infty} q_{i'}^\alpha\geq \hat{q}_j^\alpha.
\end{align}
Hence we have $\sum_i\hat{q}_i^\alpha\leq \sum_iq_i^\alpha$.
\end{IEEEproof}

\subsection{Proof of Theorem \ref{prop:Q}}\label{sec:proof_prop_Q_main}

Given $Q(x,y)\in\cB^\varepsilon(P_{XY})$, we can see that $\gamma_{xy}\eqtri Q(x,y)/P_{XY}(x,y)$ satisfies (i) $0\leq\gamma_{xy}\leq 1$ for each $(x,y)\in\cX\times\cY$ and (ii) $\sum_{x,y}\gamma_{xy}P_{XY}(x,y)\geq 1-\varepsilon$.
Hence, we can rewrite \eqref{eq:rewrite} as
\begin{align}
 H_\alpha^\varepsilon(X|Y)&=
\inf
\frac{\alpha}{1-\alpha}\log \sum_{y\in\cY}P_Y(y)\left[
\sum_{x\in\cX}\left[\frac{\gamma_{xy}P_{XY}(x,y)}{P_Y(y)}\right]^\alpha\right]^{1/\alpha}\\
&=
\inf
\frac{\alpha}{1-\alpha}\log \sum_{y\in\cY}P_Y(y)\left[
\sum_{x\in\cX}\left[\gamma_{xy}P_{X|Y}(x|y)\right]^\alpha\right]^{1/\alpha}
\label{eq:rewrite2}
\end{align}
where $\inf$ is taken over all $(\gamma_{xy})_{(x,y)\in\cX\times\cY}$ such that  (i) $0\leq\gamma_{xy}\leq 1$ for each $(x,y)\in\cX\times\cY$ and (ii) $\sum_{x,y}\gamma_{xy}P_{XY}(x,y)\geq 1-\varepsilon$.

Now, suppose that $(\gamma_{xy})_{(x,y)\in\cX\times\cY}$ satisfies (i) and (ii) above. Then, since $\sum_x\gamma_{xy}P_{XY}(x,y)\leq \sum_xP_{XY}(x,y)=P_Y(y)$, we can see that $\gamma_y\eqtri\sum_x\gamma_{xy}P_{XY}(x,y)/P_Y(y)=\sum_x\gamma_{xy}P_{X|Y}(x|y)$ satisfies (i) $0\leq\gamma_y\leq 1$ for all $y\in\cY$ and (ii) $\sum_{y}\gamma_yP_Y(y)=\sum_{x,y}\gamma_{xy}P_{XY}(x,y)\geq 1-\varepsilon$.
In other words, $\varepsilon_y\eqtri 1-\gamma_y$ and $Q_{X|Y}(x|y)\eqtri \gamma_{xy}P_{X|Y}(x|y)$ satisfy 
(i) $0\leq\varepsilon_y\leq 1$, (ii) $\sum_{y}\varepsilon_yP_Y(y)\leq \varepsilon$, (iii) $0\leq Q_{X|Y}(x|y)\leq P_{X|Y}(x|y)$, and (iv) $\sum_{x}Q_{X|Y}(x|y)= 1-\varepsilon_y$.
On the other hand, given $(\varepsilon_y)_{y\in\cY}$ and $Q_{X|Y}$ satisfying (i)--(iv), we can determine the corresponding $(\gamma_{xy})_{(x,y)\in\cX\times\cY}$. This observation brings another representation of \eqref{eq:rewrite2}:
\begin{align}
 H_\alpha^\varepsilon(X|Y)&=
\inf_{(\varepsilon_y)_{y\in\cY}\in\cE(\varepsilon)}
\frac{\alpha}{1-\alpha}\log \sum_{y\in\cY}P_Y(y)\left[
\inf_{Q_{X|Y}\in\cB_0^{\varepsilon_y}(P_{X|Y},y)}\sum_{x\in\cX}\left[Q_{X|Y}(x|y)\right]^\alpha\right]^{1/\alpha}
\end{align}
where (i) $\cE(\varepsilon)$ is the set of $(\varepsilon_y)_{y\in\cY}$ satisfying $0\leq \varepsilon_y\leq 1$ and
$\sum_y\varepsilon_yP_Y(y)\leq\varepsilon$, and (ii) $\cB_0^{\varepsilon_y}(P_{X|Y},y)$ is the set of non-negative functions $Q_{X|Y}(\cdot|y)$ on $\cX$ such that $Q_{X|Y}(x|y)\leq P_{X|Y}(x|y)$  and $\sum_xQ_{X|Y}(x|y)= 1-\varepsilon_y$.
From Lemma \ref{lemma:Schur2}, we can see that, for all $y\in\cY$,
\begin{align}
\sum_{i=1}^{i_y^*(\varepsilon_y)}\left[Q_{\varepsilon_y}^*(x_y^i|y)\right]^\alpha=
 \inf_{Q_{X|Y}\in\cB_0^{\varepsilon_y}(P_{X|Y},y)}\sum_{x\in\cX}\left[Q_{X|Y}(x|y)\right]^\alpha
\end{align}
where $i_y^*(\varepsilon_y)$ and $Q_{\varepsilon_y}^*$ are defined as in \eqref{eq:def:istar} and \eqref{eq:def:Qstar}.
Further, for any $(\varepsilon_y)_{y\in\cY}\in\cE(\varepsilon)$, there exists $(\hat{\varepsilon}_y)_{y\in\cY}\in\cE_0(\varepsilon)$ such that $\varepsilon_y\leq \hat{\varepsilon}_y$ and thus
\begin{align}
\sum_{i=1}^{i_y^*(\hat{\varepsilon}_y)}\left[Q_{\hat{\varepsilon}_y}^*(x_y^i|y)\right]^\alpha
\leq \sum_{i=1}^{i_y^*(\varepsilon_y)}\left[Q_{\varepsilon_y}^*(x_y^i|y)\right]^\alpha,\quad y\in\cY.
\end{align}
Hence we have \eqref{eq:prop:Q}.
\qed

\section{Proof of \eqref{eq:nonnegativity}}\label{sec:proof_eq_nonnegativity}
From \eqref{eq:rewrite}, we have
\begin{align}
 H_\alpha^\varepsilon(X^n|Y^n)&=
\inf_{Q_n\in\cB^\varepsilon(P_{X^nY^n})}
\frac{\alpha}{1-\alpha}\log \sum_{y^n\in\cY^n}P_{Y^n}(y^n)\left[
\sum_{x^n\in\cX^n}\left[\frac{Q_n(x^n,y^n)}{P_{Y^n}(y^n)}\right]^\alpha\right]^{1/\alpha}.
\end{align}
Since 
\begin{align}
\frac{Q_n(x^n,y^n)}{P_{Y^n}(y^n)}\leq \frac{P_{X^nY^n}(x^n,y^n)}{P_{Y^n}(y^n)}=P_{X^n|Y^n}(x^n|y^n)\leq 1,
\end{align}
we have $Q_n(x^n,y^n)/P_{Y^n}(y^n)\leq [Q_n(x^n,y^n)/P_{Y^n}(y^n)]^\alpha$ for $\alpha\in(0,1)$, and thus,
\begin{align}
 H_\alpha^\varepsilon(X^n|Y^n)
&\geq\inf_{Q_n\in\cB^\varepsilon(P_{X^nY^n})}\frac{\alpha}{1-\alpha}\log \sum_{y^n\in\cY^n}P_{Y^n}(y^n)\left[\sum_{x^n\in\cX^n}\frac{Q_n(x^n,y^n)}{P_{Y^n}(y^n)}\right]^{1/\alpha}\\
&\stackrel{\text{(a)}}{\geq}\inf_{Q_n\in\cB^\varepsilon(P_{X^nY^n})}\frac{\alpha}{1-\alpha}\log \left[\sum_{y^n\in\cY^n}\sum_{x^n\in\cX^n}Q_n(x^n,y^n)\right]^{1/\alpha}\\
&=
\inf_{Q_n\in\cB^\varepsilon(P_{X^nY^n})}\frac{1}{1-\alpha}\log \left[\sum_{y^n\in\cY^n}\sum_{x^n\in\cX^n}Q_n(x^n,y^n)\right]\\
&\geq\frac{1}{1-\alpha}\log (1-\varepsilon)
\end{align}
where (a) follows from Jensen's inequality. Hence 
\begin{align}
 \frac{1}{n}H_\alpha^\varepsilon(X^n|Y^n)\geq \frac{1}{n(1-\alpha)}\log (1-\varepsilon).
\end{align}
Taking the inferior limit of both sides, we have \eqref{eq:nonnegativity}.
\qed

\section{Proof of Theorem \ref{thm:mixture}}\label{sec:proof_thm_mixture}

Before proving the theorem, we give several lemmas in Subsection \ref{sec:proof_thm_mixture_Lemmas}. The proof of the theorem is given in Subsection \ref{sec:proof_thm_mixture_main}.

\subsection{Lemmas}\label{sec:proof_thm_mixture_Lemmas}

\begin{lemma}
\label{lemma:tmp1_proof_thm_common}
Fix $\zeta>0$ arbitrarily. Then, there exists an integer $n_0$ so that for all $n\geq n_0$ and all $i=1,2,\dots,m$,
\begin{align}
 \Pr\left\{\frac{1}{n}\log\frac{1}{P_{X^n|Y^n}(X^n|Y^n)}\geq H(X_i|Y_i)-\zeta\right\}\geq A_{i+1}-\frac{\zeta}{2}.
\end{align}
\end{lemma}

\begin{IEEEproof}
For each $k=1,2,\dots,m$, let
\begin{align}
 \cS_k^n\eqtri\left\{(x^n,y^n):\frac{1}{n}\log\frac{1}{P_{X_k^nY_k^n}(x^n|y^n)}\geq H(X_k|Y_k)-\frac{\zeta}{2}\right\}.
\end{align}
Since i.i.d.~sources satisfy the AEP \cite{Cover2}, we can choose $n_1$ such that
\begin{align}
 \sum_{(x^n,y^n)\in\cS_k^n}P_{X_k^nY_k^n}(x^n,y^n)\geq 1-\frac{\zeta}{4},\quad\forall n\geq n_1,\forall k=1,2,\dots,m.
\end{align}
Moreover, we can choose $n_0\geq n_1$ so that
\begin{align}
 -\frac{1}{n}\log\frac{\zeta}{4}\leq\frac{\zeta}{2},\quad\forall n\geq n_0.
\end{align}

Then, for all $n\geq n_0$, any $i=1,2,\dots,m$, and any $k=1,2,\dots,i$,
\begin{align}
 \tcS_i^n\eqtri\left\{(x^n,y^n):\frac{1}{n}\log\frac{1}{P_{X^n|Y^n}(x^n|y^n)}\geq H(X_i|Y_i)-\zeta\right\}
\end{align}
and
\begin{align}
 \cT_k^n\eqtri\left\{(x^n,y^n):P_{X_k^n|Y_k^n}(x^n|y^n)\leq\frac{\zeta}{4} P_{X^n|Y^n}(x^n|y^n)\right\}
\end{align}
satisfy that
\begin{align}
 \tcS_i^n\cup\cT_k^n
&\supseteq\left\{
(x^n,y^n):\frac{1}{n}\log\frac{\zeta/4}{P_{X_k^n|Y_k^n}(x^n|y^n)}\geq H(X_i|Y_i)-\zeta
\right\}\\
&=\left\{(x^n,y^n):\frac{1}{n}\log\frac{1}{P_{X_k^n|Y_k^n}(x^n|y^n)}\geq H(X_i|Y_i)-\zeta-\frac{1}{n}\log\frac{\zeta}{4}\right\}\\
&\supseteq\left\{(x^n,y^n):\frac{1}{n}\log\frac{1}{P_{X_k^n|Y_k^n}(x^n|y^n)}\geq H(X_i|Y_i)-\frac{\zeta}{2}\right\}\\
&\supseteq\cS_k^n.
\end{align}
Thus, we have
\begin{align}
 \sum_{(x^n,y^n)\in\tcS_i^n}P_{X^nY^n}(x^n,y^n)
&\geq  \sum_{k=1}^i\alpha_k\sum_{(x^n,y^n)\in\tcS_i^n}P_{X_k^nY_k^n}(x^n,y^n)\\
&\geq  \sum_{k=1}^i\alpha_k\sum_{(x^n,y^n)\in\cS_k^n}P_{X_k^nY_k^n}(x^n,y^n)-\sum_{k=1}^i\alpha_k \sum_{(x^n,y^n)\in\cT_k^n}P_{X_k^nY_k^n}(x^n,y^n)  \\
&=  \sum_{k=1}^i\alpha_k\sum_{(x^n,y^n)\in\cS_k^n}P_{X_k^nY_k^n}(x^n,y^n)-\sum_{k=1}^i\alpha_k \sum_{(x^n,y^n)\in\cT_k^n}P_{Y_k^n}(y^n)P_{X_k^n|Y_k^n}(x^n|y^n)  \\
&\geq A_{i+1}\left(1-\frac{\zeta}{4}\right)-\sum_{k=1}^i\alpha_k \sum_{(x^n,y^n)\in\cT_k^n}\frac{\zeta}{4}P_{Y_k^n}(y^n)P_{X^n|Y^n}(x^n|y^n)  \\
&\geq A_{i+1}\left(1-\frac{\zeta}{4}\right)-\frac{\zeta}{4}  \\
&\geq A_{i+1}-\frac{\zeta}{2}.
\end{align}
\end{IEEEproof}

\begin{lemma}
\label{lemma:tmp2_proof_thm_common}
Fix $\zeta>0$ arbitrarily. Then, there exists an integer $n_0$ so that for all $n\geq n_0$ and all $i=1,2,\dots,m$,
\begin{align}
 \Pr\left\{\frac{1}{n}\log\frac{1}{P_{X^nY^n}(X^n|Y^n)}\leq H(X_i|Y_i)+\zeta\right\}\geq 1-A_i-\frac{\zeta}{2}.
\end{align}
\end{lemma}

\begin{IEEEproof}
For each $k=1,2,\dots,m$, let
\begin{align}
 \cS_k^n\eqtri\left\{x^n:\frac{1}{n}\log\frac{1}{P_{X_k^n|Y_k^n}(x^n|y^n)}\leq H(X_k|Y_k)+\frac{\zeta}{3}\right\}.
\end{align}
Since i.i.d.~sources satisfy the AEP \cite{Cover2}, we can choose $n_1$ such that
\begin{align}
 \sum_{(x^n,y^n)\in\cS_k^n}P_{X_k^nY_k^n}(x^n,y^n)\geq 1-\frac{\zeta}{4},\quad\forall n\geq n_1,\forall k=1,2,\dots,m.
\end{align}
Moreover, we can choose $n_0\geq n_1$ so that
\begin{align}
 -\frac{1}{n}\log\alpha_k\leq\frac{\zeta}{3},\quad\forall n\geq n_0,\forall k=1,2,\dots,m
\end{align}
and
\begin{align}
 -\frac{1}{n}\log\frac{\zeta}{4}\leq\frac{\zeta}{3},\quad\forall n\geq n_0.
\end{align}
Then, for all $n\geq n_0$ and any $i=1,2,\dots,m$, and any $k=i,i+1,\dots,m$,
\begin{align}
 \tcS_i^n\eqtri\left\{(x^n,y^n):\frac{1}{n}\log\frac{1}{P_{X^n|Y^n}(x^n|y^n)}\leq H(X_k|Y_k)+\zeta\right\}
\end{align}
and
\begin{align}
 \cT_k^n\eqtri\cX^n\times\left\{y^n:P_{Y_k^n}(y^n)\leq\frac{\zeta}{4} P_{Y^n}(y^n)\right\}
\end{align}
satisfy that
\begin{align}
\tcS_i^n\cup\cT_k^n
&=\left\{(x^n,y^n):\frac{1}{n}\log\frac{1}{P_{X^nY^n}(x^n,y^n)}-\frac{1}{n}\log\frac{1}{P_{Y^n}(y^n)}\leq H(X_i|Y_i)+\zeta\right\}\cup\cT_k^n\\
&\supseteq\left\{(x^n,y^n):\frac{1}{n}\log\frac{1}{\alpha_kP_{X_k^nY_k^n}(x^n,y^n)}-\frac{1}{n}\log\frac{1}{P_{Y^n}(y^n)}\leq H(X_i|Y_i)+\zeta\right\}\cup\cT_k^n\\
&\supseteq\left\{(x^n,y^n):\frac{1}{n}\log\frac{1}{\alpha_kP_{X_k^nY_k^n}(x^n,y^n)}-\frac{1}{n}\log\frac{\zeta/4}{P_{Y_k^n}(y_k^n)}\leq H(X_i|Y_i)+\zeta\right\}\\
&\supseteq\left\{(x^n,y^n):\frac{1}{n}\log\frac{1}{P_{X_k^n|Y_k^n}(x^n|y^n)}\leq H(X_i|Y_i)+\frac{\zeta}{3}\right\}\\
&\supseteq\cS_k^n.
\end{align}
Thus, we have
\begin{align}
 \sum_{(x^n,y^n)\in\tcS_i^n}P_{X^nY^n}(x^n,y^n)
&\geq  \sum_{k=i}^m\alpha_k\sum_{(x^n,y^n)\in\tcS_i^n}P_{X_k^nY_k^n}(x^n,y^n)\\
&\geq  \sum_{k=i}^m\alpha_k\sum_{(x^n,y^n)\in\cS_k^n}P_{X_k^nY_k^n}(x^n,y^n)-\sum_{k=i}^m\alpha_k \sum_{(x^n,y^n)\in\cT_k^n}P_{X_k^nY_k^n}(x^n,y^n)  \\
&=  \sum_{k=i}^m\alpha_k\sum_{(x^n,y^n)\in\cS_k^n}P_{X_k^nY_k^n}(x^n,y^n)-\sum_{k=i}^m\alpha_k \sum_{(x^n,y^n)\in\cT_k^n}P_{Y_k^n}(y^n)P_{X_k^n|Y_k^n}(x^n|y^n)  \\
&\geq (1-A_i)\left(1-\frac{\zeta}{4}\right)-\sum_{k=1}^i\alpha_k \sum_{(x^n,y^n)\in\cT_k^n}\frac{\zeta}{4}P_{Y^n}(y^n)P_{X_k^n|Y_k^n}(x^n|y^n)  \\
&= (1-A_i)\left(1-\frac{\zeta}{4}\right)-\sum_{k=1}^i\alpha_k \sum_{y^n}\frac{\zeta}{4}P_{Y^n}(y^n)  \\
&\geq (1-A_i)\left(1-\frac{\zeta}{4}\right)-\frac{\zeta}{4}\\
&\geq 1-A_i-\frac{\zeta}{2}.
\end{align}
\end{IEEEproof}

\begin{lemma}
\label{lemma:tmp3_proof_thm_common}
Fix $\zeta>0$ so that $H(X_j|Y_j)-\zeta>H(X_{j+1}|Y_{j+1})+\zeta$ for all $j=1,2,\dots,m-1$.
Then, for sufficiently large $n$ and any $i=1,2,\dots,m$,
\begin{align}
 \alpha_i-\zeta\leq \Pr\left\{\abs{\frac{1}{n}\log\frac{1}{P_{X^n|Y^n}(X^n|Y^n)}- H(X_i|Y_i)}\leq\zeta\right\}\leq \alpha_i+\zeta.
\end{align}
\end{lemma}

\begin{IEEEproof}
From Lemmas \ref{lemma:tmp1_proof_thm_common} and \ref{lemma:tmp2_proof_thm_common}, we have
\begin{align}
\lefteqn{
 \Pr\left\{\abs{\frac{1}{n}\log\frac{1}{P_{X^n|Y^n}(X^n|Y^n)}- H(X_i|Y_i)}\leq\zeta\right\}
}\nonumber\\
&=  \Pr\left\{\frac{1}{n}\log\frac{1}{P_{X^n|Y^n}(X^n|Y^n)}\leq H(X_i|Y_i)+\zeta\right\}- \Pr\left\{\frac{1}{n}\log\frac{1}{P_{X^n|Y^n}(X^n|Y^n)}< H(X_i|Y_i)-\zeta\right\}\\
&\geq \{1-A_i-\zeta/2\}-\{1-(A_{i+1}-\zeta/2)\}\\
&=\alpha_i-\zeta
\end{align}
and
\begin{align}
 \lefteqn{
 \Pr\left\{\abs{\frac{1}{n}\log\frac{1}{P_{X^n|Y^n}(X^n|Y^n)}- H(X_i|Y_i)}\leq\zeta\right\}
}\nonumber\\
&=  \Pr\left\{\frac{1}{n}\log\frac{1}{P_{X^n|Y^n}(X^n|Y^n)}\leq H(X_i|Y_i)+\zeta\right\}- \Pr\left\{\frac{1}{n}\log\frac{1}{P_{X^n|Y^n}(X^n|Y^n)}< H(X_i|Y_i)-\zeta\right\}\\
&\leq  \Pr\left\{\frac{1}{n}\log\frac{1}{P_{X^n|Y^n}(X^n|Y^n)}< H(X_{i-1}|Y_{i-1})-\zeta\right\}- \Pr\left\{\frac{1}{n}\log\frac{1}{P_{X^n|Y^n}(X^n|Y^n)}\leq H(X_{i+1}|Y_{i+1})+\zeta\right\}\\
&\leq \{1-(A_i-\zeta/2)\}-\{1-A_{i+1}-\zeta/2\}\\
&=\alpha_i+\zeta.
\end{align}
\end{IEEEproof}

\subsection{Proof of Theorem \ref{thm:mixture}}\label{sec:proof_thm_mixture_main}

To prove the proposition, it is sufficient to show that, for $\varepsilon$ satisfying $A_i<\varepsilon<A_{i+1}$, 
\begin{align}
 \limsup_{n\to\infty}\frac{1}{n}H_\alpha^\varepsilon(X^n|Y^n)&\leq H(X_i|Y_i)\label{goal1:proof_thm_mixture}
\intertext{and}
 \liminf_{n\to\infty}\frac{1}{n}H_\alpha^\varepsilon(X^n|Y^n)&\geq H(X_i|Y_i).\label{goal2:proof_thm_mixture}
\end{align}

\begin{IEEEproof}
[Proof of \eqref{goal1:proof_thm_mixture}]
Fix $\zeta>0$ sufficiently small so that $H(X_j|Y_j)-\zeta>H(X_{j+1}|Y_{j+1})+\zeta$ for all $j=1,2,\dots,m-1$ and that
$A_i+m\zeta<\varepsilon$.
For $j=1,2,\dots,m$, let
\begin{align}
 \cT_n(j)\eqtri\left\{
 (x^n,y^n):\abs{
 \frac{1}{n}\log\frac{1}{P_{X^n|Y^n}(x^n|y^n)}-H(X_j|Y_j)
 }\leq \zeta
 \right\}
\label{eq:proof_thm_mixture_defT}
\end{align}
and for each $y^n\in\cY^n$ let
\begin{align}
 \cT_n(j|y^n)\eqtri\left\{x^n: (x^n,y^n)\in\cT_n(j) \right\}.
\label{eq:proof_thm_mixture_defT2}
\end{align}
Note that $\cT_n(j)\cap\cT_n(\hat{j})=\emptyset$ ($j\neq\hat{j}$).
Further, from Lemma \ref{lemma:tmp3_proof_thm_common}, we have
\begin{align}
 \Pr\left\{(X^n,Y^n)\in \bigcup_{j=i}^m\cT_n(j)\right\}
&= \sum_{j=i}^m \Pr\left\{(X^n,Y^n)\in \cT_n(j)\right\}\\
&\geq \sum_{j=i}^m \left(\alpha_j-\zeta\right)\\
&\geq 1-A_i-m\zeta\\
&\geq 1-\varepsilon. \label{eq2:proof_thm_mixture}
\end{align}
From \eqref{eq2:proof_thm_mixture}, we can see that
\begin{align}
 Q_n(x^n,y^n)\eqtri
\begin{cases}
 P_{X^nY^n}(x^n,y^n), & \text{if }(x^n,y^n)\in\bigcup_{j=i}^m\cT_n(j)\\
 0, & \text{otherwise}
\end{cases}
\end{align}
satisfies $Q_n\in\cB^\varepsilon(P_{X^nY^n})$. Thus, from the definition of $r_\alpha^\varepsilon(X^n|Y^n)$,
\begin{align}
 r_\alpha^\varepsilon(X^n|Y^n)
&\leq\sum_{y^n\in\cY^n}\left[\sum_{x^n\in\cX^n}[Q_n(x^n,y^n)]^\alpha\right]^{1/\alpha}\\
&=\sum_{y^n\in\cY^n}\left[\sum_{j=i}^m\sum_{x^n\in\cT_n(j|y^n)}[P_{X^nY^n}(x^n,y^n)]^\alpha\right]^{1/\alpha}\\
&=\sum_{y^n\in\cY^n}P_{Y^n}(y^n)\left[\sum_{j=i}^m\sum_{x^n\in\cT_n(j|y^n)}[P_{X^n|Y^n}(x^n|y^n)]^\alpha\right]^{1/\alpha}\\
&\leq \sum_{y^n\in\cY^n}P_{Y^n}(y^n)\left[\sum_{j=i}^m\abs{\cT_n(j|y^n)}\exp\{-\alpha n(H(X_j|Y_j)-\zeta)\}\right]^{1/\alpha}\\
&\leq \sum_{y^n\in\cY^n}P_{Y^n}(y^n)\left[\sum_{j=i}^m\exp\{n(H(X_j|Y_j)+\zeta)\}\exp\{-\alpha n(H(X_j|Y_j)-\zeta)\}\right]^{1/\alpha}\\
&= \sum_{y^n\in\cY^n}P_{Y^n}(y^n)\left[\sum_{j=i}^m\exp\{n[(1-\alpha)H(X_j|Y_j)+(1+\alpha)\zeta]\}\right]^{1/\alpha}\\
&\leq \sum_{y^n\in\cY^n}P_{Y^n}(y^n)\left[m\exp\{n[(1-\alpha)H(X_i|Y_i)+(1+\alpha)\zeta]\}\right]^{1/\alpha}\\
&= m^{1/\alpha}\exp\left\{n\left[\frac{1-\alpha}{\alpha}H(X_i|Y_i)+(1+\alpha)\zeta/\alpha\right]\right\}.
\end{align}
Hence, we have
\begin{align}
 \frac{1}{n}H_\alpha^\varepsilon(P_{X^n})\leq H(X_i|Y_i)+\frac{1+\alpha}{1-\alpha}\zeta+\frac{1}{n(1-\alpha)}\log m
\end{align}
and thus
\begin{align}
 \limsup_{n\to\infty}\frac{1}{n}H_\alpha^\varepsilon(P_{X^n})&\leq H(X_i|Y_i)+\frac{1+\alpha}{1-\alpha}\zeta.
\end{align}
Since we can choose $\zeta>0$ arbitrarily small, we have \eqref{goal1:proof_thm_mixture}.
\end{IEEEproof}

\begin{IEEEproof}
[Proof of \eqref{goal2:proof_thm_mixture}]
If $H(X_i)=0$ then \eqref{goal2:proof_thm_mixture} is apparent, since \eqref{eq:nonnegativity} holds.
So, we assume $H(X_i)>0$.
Fix $\zeta>0$ sufficiently small so that $H(X_j)-\zeta>H(X_{j+1})+\zeta$ for all $j=1,2,\dots,m-1$ and that
$A_i+6m\zeta<\varepsilon<A_{i+1}-6m\zeta$.
We assume that $n$ is sufficiently large so that $\exp\{-n[H(X_i|Y_i)-\zeta]\}\leq m\zeta$.

In this proof, we use $\bx$ (resp.~$\by$) instead of $x^n\in\cX^n$ (resp.~$y^n\in\cY^n$) to simplify the notation; $n$ should be apparent from the context. 
From Theorem \ref{prop:Q}, we can choose $(\varepsilon_{\by})_{\by\in\cY^n}\in\cE_0(\varepsilon)$, $Q_{\varepsilon_{\by}}^*(\cdot|\by)$, and $i_{\by}^*=i_{\by}^*(\varepsilon_{\by})$ such that
\begin{align}
 H_\alpha^\varepsilon(X^n|Y^n)+\zeta&\geq
\frac{\alpha}{1-\alpha}\log \sum_{\by\in\cY^n}P_{Y^n}(\by)\left[
\sum_{i=1}^{i_{\by}^*}\left[Q_{\varepsilon_{\by}}^*(\bx_{\by}^i|\by)\right]^\alpha\right]^{1/\alpha}
\label{eq:proof_goal2_recall}
\end{align}
where $\bx_{\by}^1,\bx_{\by}^2,\bx_{\by}^3,\dots$ are sorted so that
\begin{align}
 P_{X^n|Y^n}(\bx_{\by}^1|\by)\geq P_{X^n|Y^n}(\bx_{\by}^2|\by)\geq P_{X^n|Y^n}(\bx_{\by}^3|\by)\geq\cdots.
\end{align}
Let $\cA_n(\by)\eqtri\{\bx_{\by}^i:i=1,2,\dots,i_{\by}^*-1\}$ and $\cA_n^{+}(\by)\eqtri\cA_n(\by)\cup\{x_{\by}^{i_{\by}^*}\}$.
Then, letting
\begin{align}
 \cA_n^+\eqtri\bigcup_{\by\in\cY^n}\cA_n^+(\by)\times\{\by\},
\end{align}
we have
\begin{align}
 \Pr\{(X^n,Y^n)\in\cA_n^+\}\geq 1-\varepsilon.
\label{eq1_2:proof_goal2}
\end{align}

On the other hand, let us define $\cT_n(j)$ and $\cT_n(j|\by)$ 
as in \eqref{eq:proof_thm_mixture_defT} and \eqref{eq:proof_thm_mixture_defT2}. 
Then, from Lemma \ref{lemma:tmp3_proof_thm_common}, we have
\begin{align}
\Pr\left\{(X^n,Y^n)\notin \bigcup_{j=1}^{m}\cT_n(j)\right\}\leq m\zeta
\label{eq1:proof_goal2}
\end{align}
and
\begin{align}
 \Pr\left\{(X^n,Y^n)\in\bigcup_{j=i+1}^m\cT_n(j)\right\}&\leq \sum_{j=i+1}^m(\alpha_j+\zeta)\\
&\leq 1-A_{i+1}+m\zeta\\
&\leq 1-\varepsilon-5m\zeta.
\label{eq3:proof_goal2}
\end{align}
Thus, from \eqref{eq1_2:proof_goal2}, \eqref{eq1:proof_goal2}, and \eqref{eq3:proof_goal2}, we have
\begin{align}
 \Pr\left\{(X^n,Y^n)\in\cA_n^+\cap \bigcup_{j=1}^i\cT_n(j)\right\}\geq 4m\zeta.
\end{align}
In other words, 
\begin{align}
\sum_{\by\in\cY^n}P_{Y^n}(\by)\sum_{\bx\in \cA_n^+(\by)\cap \bigcup_{j=1}^i\cT_n(j|\by)}P_{X^n|Y^n}(\bx|\by)
\geq 4m\zeta.
\end{align}
Hence, letting
\begin{align}
 \cU_n\eqtri\left\{\by\in\cY^n:\sum_{\bx\in\cA_n^+(\by)\cap \bigcup_{j=1}^i\cT_n(j|\by)}P_{X^n|Y^n}(\bx|\by)\geq 2m\zeta\right\},
\end{align}
we have
\begin{align}
 \sum_{\by\notin\cU_n}P_{Y^n}(y^n)2m\zeta+ \sum_{\by\in\cU_n}P_{Y^n}(y^n)\geq 4m\zeta
\end{align}
and thus,
\begin{align}
 \sum_{\by\in\cU_n}P_{Y^n}(y^n)\geq\frac{2m\zeta}{1-2m\zeta}.
\label{eq9:proof_goal2}  
\end{align}

Now fix $\by\in\cU_n$. If $\bx_{\by}^{i_{\by}^*}\notin\bigcup_{j=1}^i\cT_n(j|\by)$ then
\begin{align}
 \sum_{\bx\in\cA_n(\by)\cap \bigcup_{j=1}^i\cT_n(j|\by)}P_{X^n|Y^n}(\bx|\by)
= \sum_{\bx\in\cA_n^+(\by)\cap \bigcup_{j=1}^i\cT_n(j|\by)}P_{X^n|Y^n}(\bx|\by)
\geq 2m\zeta\geq m\zeta.
\end{align}
If $\bx_{\by}^{i_{\by}^*}\in\cT_n(j|\by)$ for some $j=1,\dots,i$ then
\begin{align}
 P_{X^n|Y^n}(\bx_{\by}^{i_{\by}^*}|\by)\leq\exp\{-n[H(X_j|Y_j)-\zeta]\}\leq\exp\{-n[H(X_i|Y_i)-\zeta]\}\leq m\zeta
\end{align}
and thus
\begin{align}
 \sum_{\bx\in\cA_n(\by)\cap \bigcup_{j=1}^i\cT_n(j|\by)}P_{X^n|Y^n}(\bx|\by)
\geq \sum_{\bx\in\cA_n^+(\by)\cap \bigcup_{j=1}^i\cT_n(j|\by)}P_{X^n|Y^n}(\bx|\by)-m\zeta
\geq m\zeta.
\end{align}
Hence we have
\begin{align}
 \sum_{\bx\in\cA_n(\by)\cap \bigcup_{j=1}^i\cT_n(j|\by)}P_{X^n|Y^n}(\bx|\by)
\geq m\zeta,\quad \by\in\cU_n.
\end{align}
Further, let $p_j(\by)\eqtri\sum_{\bx\in\cA_n(\by)\cap\cT_n(j|\by)}P_{X^n|Y^n}(\bx|\by)$ for $j=1,2,\dots,i$.
Then, since $\cT_n(j|\by)\cap\cT_n(\hat{j}|\by)=\emptyset$ ($j\neq\hat{j}$), we have
\begin{align}
 \sum_{j=1}^ip_j(\by)\geq m\zeta,\quad \by\in\cU_n
\end{align}
and
\begin{align}
 \abs{\cA_n(\by)\cap\cT_n(j|\by)}\geq p_j(\by)\exp\{n[H(X_j|Y_j)-\zeta]\},\quad \by\in\cU_n,j=1,2,\dots,i.
\end{align}
Hence, for all $\by\in\cU_n$, 
\begin{align}
\sum_{\bx\in\cA_n(\by)\cap\bigcup_{j=1}^i\cT_n(j|\by)}[P_{X^n|Y^n}(\bx|\by)]^\alpha
&=\sum_{j=1}^i \sum_{x^n\in\cA_n(\by)\cap\cT_n(j|\by)}[P_{X^n|Y^n}(\bx|\by)]^\alpha\\
&\geq \sum_{j=1}^i \sum_{x^n\in\cA_n(\by)\cap\cT_n(j|\by)}\exp\{-\alpha n[H(X_j|Y_j)+\zeta]\}\\
&\geq \sum_{j=1}^i p_j(\by)\exp\{n[(1-\alpha)H(X_j|Y_j)-(1+\alpha)\zeta]\}\\
&\geq \sum_{j=1}^i p_j(\by)\exp\{n[(1-\alpha)H(X_i|Y_i)-(1+\alpha)\zeta]\}\\
&\geq m\zeta\exp\{n[(1-\alpha)H(X_i|Y_i)-(1+\alpha)\zeta]\}
\label{eq8:proof_goal2}  
\end{align}

Now, recall \eqref{eq:proof_goal2_recall}. From \eqref{eq9:proof_goal2} and \eqref{eq8:proof_goal2}, we have  
\begin{align}
 H_\alpha^\varepsilon(X^n|Y^n)+\zeta&\geq\frac{\alpha}{1-\alpha}\log \sum_{\by\in\cU_n}P_{Y^n}(\by)\left[\sum_{i=1}^{i_{\by}^*}\left[Q_{\varepsilon_{\by}}^*(\bx_{\by}^i|\by)\right]^\alpha\right]^{1/\alpha}\\
&\geq\frac{\alpha}{1-\alpha}\log \sum_{\by\in\cU_n}P_{Y^n}(\by)\left[\sum_{i=1}^{i_{\by}^*-1}\left[Q_{\varepsilon_{\by}}^*(\bx_{\by}^i|\by)\right]^\alpha\right]^{1/\alpha}\\
&=\frac{\alpha}{1-\alpha}\log \sum_{\by\in\cU_n}P_{Y^n}(\by)\left[\sum_{\bx\in\cA_n(\by)}\left[P_{X^n|Y^n}(\bx|\by)\right]^\alpha\right]^{1/\alpha}\\
&\geq\frac{\alpha}{1-\alpha}\log \sum_{\by\in\cU_n}P_{Y^n}(\by)\left[\sum_{\bx\in\cA_n(\by)\cap\bigcup_{j=1}^i\cT_n(j|\by)}\left[P_{X^n|Y^n}(\bx|\by)\right]^\alpha\right]^{1/\alpha}\\
&\geq\frac{\alpha}{1-\alpha}\log \sum_{\by\in\cU_n}P_{Y^n}(\by)
(m\zeta)^{1/\alpha}\exp\left\{n\left[\frac{1-\alpha}{\alpha}H(X_i|Y_i)-\frac{(1+\alpha)\zeta}{\alpha}\right]\right\}\\
&\geq\frac{\alpha}{1-\alpha}\log \frac{2(m\zeta)^{1+(1/\alpha)}}{1-2m\zeta}
\exp\left\{n\left[\frac{1-\alpha}{\alpha}H(X_i|Y_i)-\frac{(1+\alpha)\zeta}{\alpha}\right]\right\}\\
&=nH(X_i|Y_i)-n\frac{1+\alpha}{1-\alpha}\zeta+\frac{\alpha}{1-\alpha}\log \frac{2(m\zeta)^{1+(1/\alpha)}}{1-2m\zeta}
\end{align}
and thus
\begin{align}
 \liminf_{n\to\infty}\frac{1}{n}H_\alpha^\varepsilon(X^n|Y^n)\geq H(X_i|Y_i)-\frac{1+\alpha}{1-\alpha}\zeta.
\end{align}
Since we can choose $\zeta>0$ arbitrarily small, we have \eqref{goal2:proof_thm_mixture}.
\end{IEEEproof}

\section{Proof of Theorems \ref{thm:converse}, \ref{thm:direct}, and \ref{thm:general_guess}}\label{sec:proof}

Proof of Theorems \ref{thm:converse}, \ref{thm:direct} and \ref{thm:general_guess} is given in Subsections 
\ref{sec:proof_guess_converse}, \ref{sec:proof_guess_direct}, and \ref{sec:proof_general_guess}, respectively.

\subsection{Proof of Theorem \ref{thm:converse}}\label{sec:proof_guess_converse}

The proof of Theorem \ref{thm:converse} is essentially same as that of Theorem 1 of Arikan \cite{Arikan96}.
To prove Theorem \ref{thm:converse}, we use Lemma 1 of \cite{Arikan96}:

\medskip
\begin{lemma}[Lemma 1 of \cite{Arikan96}]\label{lemma1}
For nonnegative numbers  $a_i$ and  $q_i$ ($i=1,2,\dots,K$), and any $0<\lambda<1$, we have
\[
 \sum_{i=1}^Ka_iq_i\geq \left[\sum_{i=1}^K a_i^{\frac{-\lambda}{1-\lambda}}\right]^{\frac{1-\lambda}{-\lambda}}\left[\sum_{i=1}^Kq_i^\lambda\right]^{\frac{1}{\lambda}}.
\]
\end{lemma}

\bigskip
Given a strategy $G$, let 
\[
 Q(x,y)\eqtri\lambda_y(\sigma_y(x))P_{XY}(x,y),\quad (x,y)\in\cX\times\cY
\]
where $\lambda_i$ is defined as in \eqref{eq:zeta}.
From \eqref{eq:p_e}, we have 
\[
 \sum_{(x,y)\in\cX\times\cY}Q(x,y)= 1-p_e\geq 1-\varepsilon.
\]
Further, it is apparent that $0\leq\lambda_y(\sigma_y(x))\leq 1$ for all $(x,y)\in\cX\times\cY$, and hence
\begin{align}
 Q\in\cB^\varepsilon(P_{XY}). 
\label{eq:proof_converse}
\end{align}

Fix $y\in\cY$.
Letting $q_i=Q(\sigma_y^{-1}(i),y)/P_Y(y)$, $a_i=i^\rho$, and $\lambda=1/(1+\rho)$ in Lemma \ref{lemma1}, we have
\begin{align}
\sum_{i=1}^K\frac{Q(\sigma_y^{-1}(i),y)}{P_Y(y)}i^\rho
&\geq \left[\sum_{i=1}^Ki\right]^{-\rho}\left[\sum_{i=1}^K\left[\frac{Q(\sigma_y^{-1}(i),y)}{P_Y(y)}\right]^{\frac{1}{1+\rho}}\right]^{1+\rho}
\end{align}
and thus
\begin{align}
\sum_{x\in\cX}\lambda_y(\sigma_y(x))P_{XY}(x,y) \sigma_y(x)^\rho
&\geq \left[\sum_{i=1}^Ki\right]^{-\rho}\left[\sum_{x\in\cX}Q(x,y)^{\frac{1}{1+\rho}}\right]^{1+\rho}.
\end{align}
Taking the summation over all $y\in\cY$ and recalling \eqref{eq:barC}, we have
\begin{align}
\bar{C}_\rho
&\geq \left[\sum_{i=1}^Ki\right]^{-\rho}\sum_{y\in\cY}\left[\sum_{x\in\cX}Q(x,y)^{\frac{1}{1+\rho}}\right]^{1+\rho}.
\end{align}
Further, from the fact that
\begin{align}
 \sum_{i=1}^K\frac{1}{i}\leq 1+\log K,
\end{align}
we have
\begin{align}
 \bar{C}_\rho\geq (1+\log K)^{-\rho}\sum_{y\in\cY}\left[\sum_{x\in\cX}Q(x,y)^{\frac{1}{1+\rho}}\right]^{1+\rho}.
\end{align}
Hence, from the definition of $H_{1/(1+\rho)}^\varepsilon(X|Y)$ and \eqref{eq:proof_converse}, we have \eqref{eq:converse}.\qed

\subsection{Proof of Theorem \ref{thm:direct}}\label{sec:proof_guess_direct}
To prove Theorem \ref{thm:direct}, we construct a strategy $G$ as follows.
Recall Theorem \ref{prop:Q}: for any $\zeta>0$, we can choose $(\varepsilon_y)_{y\in\cY}\in\cE_0(\varepsilon)$, $i_y^*=i_y^*(\varepsilon_y)$, and $Q_{\varepsilon_y}^*$ satisfying 
\begin{align}
 \rho H_{\frac{1}{1+\rho}}^\varepsilon(X|Y)+\gamma&=
\log \sum_{y\in\cY}P_Y(y)\left[
\sum_{i=1}^{i_y^*}\left[Q_{\varepsilon_y}^*(x_y^i|y)\right]^{\frac{1}{1+\rho}}\right]^{1+\rho}.
\label{eq:Koga}
\end{align}
By using this notation, for each $y\in\cY$, let us define $\sigma_y$ and $\pi_y$ so that
\begin{align}
 \sigma_y(x_y^i)=i
\end{align}
and 
\begin{align}
 \pi_i\eqtri
\begin{cases}
 0 &: i< i_y^*,\\
1-\frac{Q^*(\sigma_y^{-1}(i_y^*)|y)}{P_{X|Y}(\sigma_y^{-1}(i_y^*)|y)} &: i=i_y^*,\\
 1 &: i> i_y^*.
\end{cases}
\end{align}
From the definition of $Q_{\varepsilon_y}^*$ in \eqref{eq:def:Qstar} and the definition of $\lambda_i$ in \eqref{eq:zeta},
we can see that 
\begin{align}
Q_{\varepsilon_y}^*(\sigma_y^{-1}(i)|y)=\lambda_y(i)P_{X|Y}(\sigma_y^{-1}(i)|y),\quad y\in\cY, i=1,2,\dots,K.
\end{align}

Hence, we can upper bound $\bar{C}_\rho=\bar{C}_\rho(G|X,Y)$ as follows.\footnote{The following step 
is essentially same as that of Proposition 4 of Arikan \cite{Arikan96}.}
For each $y\in\cY$, 
\begin{align}
 \sigma_y(x)&=\sum_{\tx:\sigma_y(\tx)\leq \sigma_y(x)}1\\
&\leq \sum_{\tx:\sigma_y(\tx)\leq \sigma_y(x)}\left[
\frac{Q_{\varepsilon_y}^*(\tx|y)}{Q_{\varepsilon_y}^*(x|y)}
\right]^{\frac{1}{1+\rho}}\\
&\leq \sum_{\tx\in\cX}\left[
\frac{Q_{\varepsilon_y}^*(\tx|y)}{Q_{\varepsilon_y}^*(x|y)}
\right]^{\frac{1}{1+\rho}}\\
&= \sum_{j=1}^K\left[
\frac{Q_{\varepsilon_y}^*(x_y^j|y)}{Q_{\varepsilon_y}^*(x|y)}
\right]^{\frac{1}{1+\rho}}\\
&= \sum_{j=1}^{i_y^*}\left[
\frac{Q_{\varepsilon_y}^*(x_y^j|y)}{Q_{\varepsilon_y}^*(x|y)}
\right]^{\frac{1}{1+\rho}}
\end{align}
and thus, for each $i$,
\begin{align}
 i&=\sigma_y(x_y^i)\\
&\leq \sum_{j=1}^{i_y^*}\left[
\frac{Q_{\varepsilon_y}^*(x_y^j|y)}{Q_{\varepsilon_y}^*(x_y^i|y)}
\right]^{\frac{1}{1+\rho}}.
\end{align}
Hence, we have
\begin{align}
 \bar{C}_\rho
&=\sum_{y\in\cY}P_Y(y)\left\{
\sum_{i=1}^K\lambda_y(i)P_{X|Y}(\sigma_y^{-1}(i)|y)i^\rho
\right\}\\
&=\sum_{y\in\cY}P_Y(y)\left\{\sum_{i=1}^{i_y^*}Q^*(x_y^i|y)i^\rho\right\}\\
&\leq \sum_{y\in\cY}P_Y(y)\left\{\sum_{i=1}^{i_y^*}Q^*(x_y^i|y)
\left[
\sum_{j=1}^{i_y^*}\left[
\frac{Q_{\varepsilon_y}^*(x_y^j|y)}{Q_{\varepsilon_y}^*(x_y^i|y)}
\right]^{\frac{1}{1+\rho}}
\right]
^\rho\right\}\\
&= \sum_{y\in\cY}P_Y(y)\left\{
\left[
\sum_{i=1}^{i_y^*}Q^*(x_y^i|y)
\right]
\times
\left[
\sum_{j=1}^{i_y^*}\left[
Q_{\varepsilon_y}^*(x_y^j|y)
\right]^{\frac{1}{1+\rho}}
\right]
^\rho\right\}\\
&= \sum_{y\in\cY}P_Y(y)
\left[
\sum_{i=1}^{i_y^*}\left[
Q_{\varepsilon_y}^*(x_y^i|y)
\right]^{\frac{1}{1+\rho}}
\right]
^{1+\rho}\\
&=\exp\left\{\rho H_{\frac{1}{1+\rho}}^\varepsilon(X|Y)+\gamma\right\}
\end{align}
where the last equality follows from \eqref{eq:Koga}. 
Since we can choose $\zeta>0$ arbitrarily small, we have \eqref{eq:direct}.\qed

\subsection{Proof of Theorem \ref{thm:general_guess}}\label{sec:proof_general_guess}

\begin{IEEEproof}
[Direct Part]
Fix $\delta>0$ so that $0<\varepsilon+\delta<1$. From Theorem \ref{thm:direct}, there exists $\{G_n\}_{n=1}^\infty$ such that, for $n=1,2,\dots,$
\begin{align}
 \error(G_n|X^n,Y^n)\leq\varepsilon+\delta
\end{align}
and
\begin{align}
\bar{C}_\rho(G_n|X^n,Y^n)
&\leq 
\exp\left\{\rho H_{1/(1+\rho)}^{\varepsilon+\delta}(X^n|Y^n)\right\}.
\end{align}
Hence, 
\begin{align}
 \limsup_{n\to\infty}\frac{1}{n}\log \bar{C}_\rho(G_n|X^n,Y^n)
&\leq \rho \limsup_{n\to\infty}\frac{1}{n}H_{1/(1+\rho)}^{\varepsilon+\delta}(X^n|Y^n).
\end{align}
By using the \emph{diagonal line argument} (see \cite{Han-spectrum}), we can conclude that $\rho H_{1/(1+\rho)}^\varepsilon(\bX|\bY)$ is $\varepsilon$-achievable.
\end{IEEEproof}

\begin{IEEEproof}
[Converse Part]
Suppose that $\expguess$ is $\varepsilon$-achievable and fix $\delta>0$ arbitrarily.
Then there exists $\{G_n\}_{n=1}^\infty$ such that, for sufficiently large $n$,
\begin{align}
 \error(G_n|X^n,Y^n)\leq \varepsilon+\delta
\label{eq4:proof_prop_gen_guess}
\end{align}
and
\begin{align}
 \limsup_{n\to\infty}\frac{1}{n}\log \bar{C}_\rho(\code_n|X^n,Y^n)\leq \expguess.
\label{eq2:proof_prop_gen_guess}
\end{align}
On the other hand, from Theorem \ref{thm:converse}, for sufficiently large $n$ such that 
\eqref{eq4:proof_prop_gen_guess}
holds, 
\begin{align}
\bar{C}_\rho(G_n|X^n,Y^n)
\geq (1+\log K)^{-\rho}\exp\left\{\rho H_{1/(1+\rho)}^{\varepsilon+\delta}(X^n|Y^n)\right\}.
\label{eq3:proof_prop_gen_guess}
\end{align}
Combining \eqref{eq2:proof_prop_gen_guess} with \eqref{eq3:proof_prop_gen_guess}, we have
\begin{align}
 \expguess\geq \rho\limsup_{n\to\infty}\frac{1}{n}H_{1/(1+\rho)}^{\varepsilon+\delta}(X^n|Y^n).
\end{align}
Since $\delta>0$ is arbitrary, letting $\delta\downarrow 0$, we have
$\expguess\geq \rho H_{1/(1+\rho)}^{\varepsilon}(\bX|\bY)$.
\end{IEEEproof}

\section{Proof of Theorems \ref{thm:main_converse}, \ref{thm:main_direct}, and \ref{thm:gen}}\label{sec:proof_sourcecoding}

Proof of Theorems \ref{thm:main_converse}, \ref{thm:main_direct}, and \ref{thm:gen} is given in Subsections 
\ref{sec:proof_converse}, \ref{sec:proof_direct}, and \ref{sec:proof_thm_gen}, respectively.

\subsection{Proof of Theorem \ref{thm:main_converse}}\label{sec:proof_converse}

Fix a code $\code$ such that $\error(\code)\leq 1-\varepsilon$.

Fix $y\in\cY$.
Recall that we allow $\varphi_y$ to be stochastic. Let $W_y(c|x)\eqtri W_{\varphi_y}(c|x)$ be the probability such that $x\in\cX$ is mapped to $c\in\cC$ by $\varphi_y$.
Let
\begin{align}
 \Gamma_y(x)\eqtri\left\{c\in\cC_y:W_y(c|x)>0,x=\psi_y(c)\right\}
\end{align}
and
\begin{align}
 \gamma_{xy}\eqtri\sum_{c\in\Gamma_y(x)}W_y(c|x).
\end{align}
Note that, since $\error(\code)\leq \varepsilon$, we have
\begin{align}
 \sum_{(x,y)\in\cX\times\cY}\gamma_{xy}P_{XY}(x,y)\geq 1-\varepsilon.
\label{eq3:converse}
\end{align}

Further, using Jensen's inequality, we have
\begin{align}
\sum_{c\in\cC_y}W_y(c|x)\exp\{\rho\funcabs{c}\log 2\}
&\geq \sum_{c\in\Gamma_y(x)}W_y(c|x)\exp\{\rho\funcabs{c}\log 2\}\\
&\geq \gamma_{xy}\exp\left\{\rho\sum_{c\in\Gamma_y(x)}\frac{W_y(c|x)}{\gamma_{xy}}\funcabs{c}\log 2\right\}\\
&\geq \gamma_{xy}\exp\left\{\rho\sum_{c\in\Gamma_y(x)}\frac{W_y(c|x)}{\gamma_{xy}}\bar\ell(x|y)\right\}\\
&= \gamma_{xy}\exp\left\{\rho\bar\ell(x|y)\right\}
\label{eq2:converse}
\end{align}
where
\begin{align}
 \bar\ell(x|y)\eqtri\min_{c\in\Gamma_y(x)}\funcabs{c}\log 2.
\end{align}

Substituting \eqref{eq2:converse} into \eqref{eq:def_moment}, we have
\begin{align}
M_\rho(\code|X,Y)\geq \sum_{(x,y)\in\cX\times\cY}\gamma_{xy}P_{XY}(x,y)\exp\left\{\rho\bar\ell(x|y)\right\}.
\label{eq4:converse}
\end{align}
Let $Q(x,y)=\gamma_{xy}P_{XY}(x,y)$. Then, from \eqref{eq3:converse}, we have $Q\in\cB^\varepsilon(P_{XY})$.
Further, let $Q(x|y)\eqtri Q(x,y)/P_Y(y)$ and $\cA_y\eqtri \{x:Q(x|y)>0\}$ for each $y\in\cY$. Then, \eqref{eq4:converse} can be written as
\begin{align}
M_\rho\geq \sum_{y\in\cY}P_Y(y)\sum_{x\in\cA_x}Q(x|y)\exp\left\{\rho\bar\ell(x|y)\right\}.
\label{eq6:converse}
\end{align}

Now, fix $y\in\cY$. From the definition of the set $\Gamma_y(x)$, we can see that
$\Gamma_y(x)\cap\Gamma_y(x')=\emptyset$ for all $x,x'\in\cA_y$ such that $x\neq x'$.
Thus, from the prefix condition, we have
\begin{align}
 \sum_{x\in\cA_y}\exp\{-\bar\ell(x|y)\}\leq 1.
\label{eq5:converse}
\end{align}
Then, let us consider the problem of minimizing
$\sum_{x\in\cA_y}Q(x|y)\exp\left\{\rho\bar\ell(x|y)\right\}$ subject to \eqref{eq5:converse}.
As shown in Example 1 in Section 3 of \cite{Merhav_arXiv2011}, the minimum is achieved by
\begin{align}
 \bar\ell(x|y)=-\log\frac{[Q(x|y)]^{1/(1+\rho)}}{\sum_{x'\in\cA_y}[Q(x'|y)]^{1/(1+\rho)}},\quad x\in\cA.
\end{align}

Applying the above argument for each $y\in\cY$, we can rewrite \eqref{eq6:converse} as
\begin{align}
M_\rho&\geq \sum_{y\in\cY}P_Y(y)\sum_{x\in\cA_x}Q(x|y)\exp\left\{\rho\bar\ell(x|y)\right\}\\
&\geq \sum_{y\in\cY}P_Y(y)\left\{
\sum_{x\in\cA_y}Q(x|y)\exp\left[-\rho
\log\frac{[Q(x|y)]^{1/(1+\rho)}}{\sum_{x'\in\cA_y}[Q(x'|y)]^{1/(1+\rho)}}
\right]\right\}\\
&= \sum_{y\in\cY}P_Y(y)\left\{
\sum_{x\in\cA_y}Q(x|y)\exp\left[-\rho
\log\frac{[Q(x,y)]^{1/(1+\rho)}}{\sum_{x'\in\cA_y}[Q(x',y)]^{1/(1+\rho)}}
\right]\right\}\\
&= 
\sum_{y\in\cY}\sum_{x\in\cA_x}Q(x,y)\exp\left[-\rho
\log\frac{[Q(x,y)]^{1/(1+\rho)}}{\sum_{x'\in\cA_y}[Q(x',y)]^{1/(1+\rho)}}
\right]\\
&= 
\sum_{y\in\cY}\sum_{x\in\cA_x}Q(x,y)\left[
\frac{[Q(x,y)]^{1/(1+\rho)}}{\sum_{x'\in\cA_y}[Q(x',y)]^{1/(1+\rho)}}
\right]^{-\rho}\\
&= 
\sum_{y\in\cY}\sum_{x\in\cA_y}[Q(x,y)]^{1/(1+\rho)}
\left[\sum_{x'\in\cA_y}[Q(x',y)]^{1/(1+\rho)}\right]^\rho\\
&= 
\sum_{y\in\cY}\left[
\sum_{x\in\cA_y}[Q(x,y)]^{1/(1+\rho)}
\right]^{1+\rho}\\
&= 
\sum_{y\in\cY}\left[
\sum_{x\in\cX}[Q(x,y)]^{1/(1+\rho)}
\right]^{1+\rho}\\
&\geq r_{1/(1+\rho)}^\varepsilon(X|Y)
\end{align}
where the last inequality follows from the fact $Q\in\cB^\varepsilon(P)$.

By the definition of the conditional smooth R\'enyi entropy, we have \eqref{eq:thm_main_converse}.
\qed

\subsection{Proof of Theorem \ref{thm:main_direct}}\label{sec:proof_direct}

Fix $\zeta>0$ arbitrarily and choose $Q\in\cB^\varepsilon(P_{XY})$ so that
\begin{align}
 \log\sum_{y\in\cY}\left[\sum_{x\in\cX}[Q(x,y)]^{1/(1+\rho)}\right]^{1+\rho}\leq \rho H_{1/(1+\rho)}^\varepsilon(X|Y) +\zeta.
\end{align}
Letting $Q(x|y)\eqtri Q(x,y)/P_Y(y)$, we have
\begin{align}
 \log\sum_{y\in\cY}P_Y(y)\left[\sum_{x\in\cX}[Q(x|y)]^{1/(1+\rho)}\right]^{1+\rho}\leq \rho H_{1/(1+\rho)}^\varepsilon(X|Y) +\zeta.
\label{eq1:proof_direct}
\end{align}

Fix $y\in\cY$. Let $\cA_y\eqtri\{x\in\cX:Q(x|y)>0\}$ and
\begin{align}
 \tQl(x|y)&\eqtri\frac{[Q(x|y)]^{1/(1+\rho)}}{\sum_{x'\in\cA}[Q(x'|y)]^{1/(1+\rho)}}.
\end{align}
Since
\begin{align}
 \sum_{x\in\cA_x}2^{-\{-\log_2 \tQl(x|y)\}}\leq 1
\end{align}
holds, we can construct a variable-length code $(\hat\varphi_y,\hat\psi_y,\hat\cC_y)$ such that (i) $\hat\cC_y\eqtri\{\hat\varphi_y(x):x\in\cA_y\}$ is prefix free, (ii)
$\hat\varphi_y\colon\cA_y\to\hat\cC_y$ satisfies
\begin{align}
 \funcabs{\hat\varphi_y(x)}=\lceil-\log_2 \tQl(x|y)\rceil,
\end{align}
and, (iii) $\hat\varphi_y$ and $\hat\psi_y\colon\hat\cC_y\to\cA_y$ satisfy $x=\psi_y(\varphi_y(x))$ for all  $x\in\cA_y$; e.g.~we can use the Shannon code for the distribution $\tQl(x|y)$.
Further, for each $x\in\cX$, let $\gamma_{xy}=Q(x|y)/P_{X|Y}(x|y)$. Note that $0\leq\gamma_{xy}\leq 1$ and $\gamma_{xy}=0$ for all $x\notin\cA_y$.
Then, we construct a stochastic encoder for $X$ given $y$ as follows:
\begin{align}
 \varphi_y(x)=
\begin{cases}
 0\circ\hat\varphi_y(x) & \text{with probability }\gamma_{xy}\\
 1 & \text{with probability }1-\gamma_{xy}
\end{cases}
\end{align}
where $\circ$ denotes the concatenation. 
That is, $x$ is encoded to ``0'' following $\hat\varphi_y(x)$ with probability $\gamma_{xy}$, and ``1'' with probability $1-\gamma_{xy}$.
We can construct the corresponding decoder $\psi_y$ so that $x=\psi_y(\varphi_y(x))$ for all $x\in\cA_y$ if $x$ is encoded to $0\circ\hat\varphi_y(x)$.
The length function $\ell(x|y)=\funcabs{\varphi_y(x)}\log 2$ satisfies that, if $x$ is encoded to ``0'' following $\hat\varphi_y(x)$, 
\begin{align}
\ell(x|y)&\leq -\log \tQl(x|y)+2\log 2
\end{align}
and otherwise $\ell(x|y)=\log 2$.

By applying the above argument  for each $y\in\cY$, we have
\begin{align}
M_\rho&=\Ex_{P_{XY}}\left[\exp\{\rho\ell(X|Y)\}\right]\\
&\leq \sum_{y\in\cY}P_Y(y)\sum_{x\in\cX}P_{X|Y}(x|y)\gamma_{xy}\exp\left\{\rho[-\log \tQl(x|y)+2\log 2]\right\}\nonumber\\
&\qquad+\sum_{y\in\cY}P_Y(y)\sum_{x\in\cX}P_{X|Y}(x|y)(1-\gamma_{xy})\exp\{\rho\log 2\}\\
&\stackrel{\text{(a)}}{\leq} 2^{2\rho}\sum_{y\in\cY}P_Y(y)\sum_{x\in\cA_y}Q(x|y)\exp\left\{-\rho\log \tQl(x|y)\right\}
+\varepsilon 2^{\rho}\\
&= 2^{2\rho}\sum_{y\in\cY}P_Y(y)\left\{\sum_{x\in\cA_y}[Q(x|y)]^{1/(1+\rho)}\right\}^{(1+\rho)}
+\varepsilon 2^{\rho}\\
&\stackrel{\text{(b)}}{\leq} 2^{2\rho}\exp\left\{\rho H_{1/(1+\rho)}^\varepsilon(P)+\zeta\right\}
+\varepsilon 2^{\rho}
\end{align}
where the inequality (a) follows from the fact that, since  $Q\in\cB^\varepsilon(P)$, 
\begin{align}
 \sum_{y\in\cY}P_{Y}(y)\sum_{x\in\cX}P_{X|Y}(x|y)\gamma_{xy}=\sum_{x,y}Q(x,y)\geq 1-\varepsilon
\label{eq2:proof_direct}
\end{align}
and the inequality (b) follows from \eqref{eq1:proof_direct}.

Since we can choose $\zeta>0$ arbitrarily small, we have \eqref{eq:thm_main_direct}.
\qed

\subsection{Proof of Theorem \ref{thm:gen}}\label{sec:proof_thm_gen}
\begin{IEEEproof}
[Direct Part]
Fix $\delta>0$ so that $0<\varepsilon+\delta<1$. From Theorem \ref{thm:main_direct}, there exists $\{\code_n\}_{n=1}^\infty$ such that, for $n=1,2,\dots,$
\begin{align}
 \error(\code_n|X^n,Y^n)\leq\varepsilon+\delta
\end{align}
and
\begin{align}
M_\rho(\code_n|X^n,Y^n)
&\leq 
\max\left\{
2\times2^{2\rho}\exp\left\{\rho H_{1/(1+\rho)}^{\varepsilon+\delta}(X^n|Y^n)\right\},
2\times(\varepsilon+\delta) 2^{\rho}
\right\}.
\label{eq1:proof_prop_gen} 
\end{align}
From  \eqref{eq:nonnegativity} and \eqref{eq1:proof_prop_gen}, we can see
\begin{align}
 \limsup_{n\to\infty}\frac{1}{n}\log M_\rho(\code_n|X^n,Y^n)
&\leq \rho \limsup_{n\to\infty}\frac{1}{n}H_{1/(1+\rho)}^{\varepsilon+\delta}(X^n|Y^n).
\end{align}
By using the \emph{diagonal line argument} (see \cite{Han-spectrum}), we can conclude that $\rho H_{1/(1+\rho)}^\varepsilon(\bX|\bY)$ is $\varepsilon$-achievable.
\end{IEEEproof}

\begin{IEEEproof}
[Converse Part]
Suppose that $\expsc$ is $\varepsilon$-achievable and fix $\delta>0$ arbitrarily.
Then there exists $\{\code_n\}_{n=1}^\infty$ such that, for sufficiently large $n$,
\begin{align}
 \error(\code_n|X^n,Y^n)\leq \varepsilon+\delta
\label{eq4:proof_prop_gen}
\end{align}
and
\begin{align}
 \limsup_{n\to\infty}\frac{1}{n}\log M_\rho(\code_n|X^n,Y^n)\leq \expsc.
\label{eq2:proof_prop_gen}
\end{align}
On the other hand, from Theorem \ref{thm:main_converse}, for sufficiently large $n$ such that 
\eqref{eq4:proof_prop_gen}
holds, 
\begin{align}
M_\rho(\code_n|X^n,Y^n)
\geq \exp\left\{\rho H_{1/(1+\rho)}^{\varepsilon+\delta}(X^n|Y^n)\right\}.
\label{eq3:proof_prop_gen}
\end{align}
Combining \eqref{eq2:proof_prop_gen} with \eqref{eq3:proof_prop_gen}, we have
\begin{align}
 \expsc\geq \rho\limsup_{n\to\infty}\frac{1}{n}H_{1/(1+\rho)}^{\varepsilon+\delta}(X^n|Y^n).
\end{align}
Since $\delta>0$ is arbitrary, letting $\delta\downarrow 0$, we have
$\expsc\geq \rho H_{1/(1+\rho)}^{\varepsilon}(\bX|\bY)$.
\end{IEEEproof}




\end{document}